\pdfoutput=1
\newif\ifpreprint
\preprinttrue
\newif\ifnotpreprint
\ifpreprint
\notpreprintfalse
\else
\notpreprinttrue
\fi

\documentclass[reprint, amsmath,amssymb, nobibnotes, aps, prb, superscriptaddress, citeautoscript, nofootinbib]{revtex4-2}
\usepackage{jabbrv}
\usepackage{graphicx}% Include figure files
\usepackage{hyperref}% add hypertext capabilities
\usepackage{booktabs}
\usepackage{xcolor}
\usepackage{dcolumn}
\usepackage{rotating}
\usepackage{longtable}
\usepackage{makecell}
\makeatletter
\let\auto@bib@innerbib\@empty
\let\write@bibliographystyle\relax
\makeatother

\usepackage[utf8]{inputenc}
\usepackage[T1]{fontenc}
\usepackage{orcidlink}

\usepackage[normalem]{ulem}
\usepackage[version=4]{mhchem}
\usepackage{bm}
\usepackage{bbold}
\usepackage{braket}
\usepackage{siunitx}

\DeclareMathAlphabet\mathbfcal{OMS}{cmsy}{b}{n}

\newcommand{\editorr}[2]{%
  \expandafter\newcommand\csname #1note\endcsname[1]{%
    \textcolor{#2}{(\textbf{#1:} ##1)}}%
  \expandafter\newcommand\csname #1\endcsname[1]{%
    \textcolor{#2}{##1}}%
  \expandafter\newcommand\csname #1cancel\endcsname[1]{%
    \textcolor{#2}{\sout{##1}}}%
  \expandafter\newcommand\csname #1change\endcsname[2]{%
    \textcolor{#2}{\sout{##1} ##2}}%
  \newenvironment{#1text}{\color{#2}}{\color{black}}
}

\editorr{LB}{red}
\editorr{CC}{violet}

\newcommand\THEOSMARVEL{Theory and Simulation of Materials (THEOS), and National Centre for Computational Design and Discovery of Novel Materials (MARVEL), {\'E}cole Polytechnique F{\'e}d{\'e}rale de Lausanne, 1015 Lausanne, Switzerland}
\newcommand\PSI{Laboratory for Materials Simulations, Paul Scherrer Institut (PSI), 5232 Villigen, Switzerland}
\newcommand\BREMEN{U Bremen Excellence Chair, Bremen Center for Computational Materials 
Science, and MAPEX Center for Materials and Processes, Universit\"at Bremen, 28359 Bremen, Germany}
\newcommand\SAPIENZA{Department of Physics, Sapienza Universit\`a di Roma, Rome, Italy}
\newcommand\RENNES{Univ Rennes, ENSCR, CNRS, Institut des Sciences Chimiques de Rennes, UMR 6226, Rennes, France}
\newcommand\CAMBRIDGE{Theory of Condensed Matter, Cavendish Laboratory, University of Cambridge, Cambridge, CB3 0US, United Kingdom}

\begin{document}
\ifpreprint
\fi
% ============================= TITLE
\title{Predicting accurate solid-state thermodynamics is easier than you think}
\ifpreprint

\author{{Lorenzo Bastonero}\orcidlink{0000-0001-9374-1876}} \email{lbastone@uni-bremen.de} \affiliation{\BREMEN}
\author{{Gabriel Joalland}\orcidlink{0009-0000-1129-6685}} \affiliation{\THEOSMARVEL} \affiliation{\RENNES}
\author{{Chiara Cignarella}\orcidlink{0000-0002-0603-4389}} \affiliation{\BREMEN} \affiliation{\THEOSMARVEL}
\author{{Lorenzo Monacelli}\orcidlink{0000-0002-6381-3741}} \affiliation{\THEOSMARVEL} \affiliation{\SAPIENZA}
\author{{Nicola Marzari}\orcidlink{0000-0002-9764-0199}} \affiliation{\BREMEN} \affiliation{\THEOSMARVEL} \affiliation{\CAMBRIDGE} \affiliation{\PSI}

\else

\author[1,*]{Lorenzo Bastonero}
\author[2,3]{Gabriel Joalland}
\author[1,2]{Chiara Cignarella}
\author[2,4]{Lorenzo Monacelli}
\author[1,2,5,6]{Nicola Marzari}
\affil[1]{\BREMEN}
\affil[2]{\THEOSMARVEL}
\affil[3]{\RENNES}
\affil[4]{\SAPIENZA}
\affil[5]{\CAMBRIDGE}
\affil[6]{\PSI}
\affil[*]{lbastone@uni-bremen.de}
\fi

% ============================= ABSTRACT
% ==========================================================
\begin{abstract}
Materials underpin modern technologies, from energy harvesting, storage, and conversion to information and communication technologies. Their functionality is often governed by the interplay between competing phases, and thermodynamic behavior shapes microscopic properties and ultimately determines technological performance; for instance, the light absorption of inorganic metal-halide perovskites in solar cells. Accurately predicting crystal thermodynamics, however, remains a major challenge for computational approaches because strong anharmonic effects require extensive sampling of the potential energy surface. Here, we present an on-the-fly Bayesian framework, combined with the stochastic self-consistent harmonic approximation, for learning first-principles interatomic potentials. This approach enables the prediction of thermodynamic properties over a broad temperature range with first-principles accuracy while requiring training on only a few tens to a few hundreds of atomic configurations. A relevant finding is that the majority of the required structures are obtained from a single inexpensive dataset generation step, demonstrating that the relevant physics for accurately predicting solid-state thermodynamics is encoded in a few physically meaningful configurations. To demonstrate its power, we investigate the thermodynamic and dynamical properties of Li$_2$O, $\alpha$--CsPbI$_3$, and $\delta$--CsPbI$_3$, requiring only 44, 256, and 50 total-energy calculations, respectively. Notably, we show that this framework accurately captures the extreme thermal expansion of \ce{Li2O}, and the phase diagram of CsPbI$_3$, which explains its spontaneous degradation into the non-absorbing yellow phase, with a predicted transition temperature within 20~K of experiment. More broadly, the method presented opens an interesting route toward accelerated materials engineering under realistic conditions for a wide range of technologically relevant applications, including solid-state batteries, optoelectronic devices, and memristors.
\end{abstract}

\flushbottom
\maketitle
\thispagestyle{empty}

% ============================= MAIN CONTENT
% ==========================================================
\section{Introduction}
Materials are a cornerstone of our technological world, starting, e.g., from a green transition supported by breakthroughs in the conversion, storage, and use of clean energy. 
Materials discovery and design is therefore central to the development of more sustainable, efficient, and scalable energy technologies.
However, the desired functionalities of these materials are often controlled by atomic rearrangements occurring only at a specific temperature. 
This is the case, for example, of liquid-like ionic mobility in superionic conductors for Li-ion batteries~\cite{Wang2015, Muy2025}, optoelectronic response of halide perovskites for energy harvesting~\cite{Jiang2018}, and lattice thermal transport of thermoelectrics for waste heat management~\cite{Aseginolaza2019,Zeng2025}.
Moreover, many scientifically and technologically relevant compounds sit close to phase transitions, driven by unstable vibrational modes~\cite{Monacelli2023}, strong anharmonicity~\cite{Aseginolaza2019}, and even non-negligible nuclear quantum fluctuations in light-element systems~\cite{Errea2020}.
In these regimes, the commonly-used harmonic and quasi-harmonic approximations are inadequate to describe the thermodynamic driving forces~\cite{Errea2020, Troyan2021, Monacelli2021a, Ranalli2023}. 

Several nonperturbative frameworks have been developed to incorporate anharmonic and quantum nuclear effects beyond perturbation theory, including self-consistent phonon approaches based on polynomial expansions of the Born--Oppenheimer energy landscape~\cite{Tadano2014,Tadano2015,Tadano2018}, self-consistent \emph{ab initio} lattice dynamics and its quantum extension~\cite{Souvatzis2008,Souvatzis2009,Roekeghem2021}, the temperature-dependent effective potential (TDEP) method~\cite{Hellman2011,Hellman2013a,Hellman2013,Knoop2024}, also in its stochastic quantum formulation~\cite{Shulumba2017}, the anharmonic special displacement method~\cite{Zacharias2020,Zacharias2023}, and the stochastic self-consistent harmonic approximation (SSCHA)~\cite{Errea2013,Bianco2017,Monacelli2018,Monacelli2021a}.
These approaches were designed around different targets, and differ in their variational or fitting objective, in how configurations are sampled, and in the treatment of the lattice degrees of freedom, higher-order force constants, as well as static and dynamical responses.
Describing thermodynamic properties, phase transitions, and constructing phase diagrams, however, require a comprehensive framework, in which free energies, structural relaxations, and vibrational response functions are all accessible on the same footing: first-order boundaries follow from the crossing of the Gibbs free energies of the competing phases, each relaxed with respect to both internal coordinates and lattice parameters, whereas continuous transitions are signalled by a vanishing eigenvalue of the free-energy Hessian, i.e., by the softening of an anharmonic vibrational mode.
We therefore adopt the SSCHA, which minimizes the free energy over a Gaussian trial nuclear density matrix and provides precisely these ingredients within a single framework~\cite{Bianco2017, Monacelli2018}.
On this basis, the SSCHA has already been used to compute thermodynamic properties and resolve finite-temperature phase boundaries in charge-density waves~\cite{Bianco2019,Diego2021,McGuinness2026}, high-pressure ice~\cite{Cherubini2024}, hydrates~\cite{Ranieri2023}, hydrides~\cite{Errea2016,Errea2020,Troyan2021}, hydrogen~\cite{Borinaga2016,Monacelli2023a}, ferroelectrics~\cite{Aseginolaza2019}, one-dimensional materials~\cite{Romanin2021, Cignarella2025}, and perovskites~\cite{Verdi2023,Ranalli2023,Monacelli2023}.
However, SSCHA calculations can be demanding for complex materials because they require many \emph{ab-initio} total-energy and force evaluations on large supercells, which represents a bottleneck for large-scale applications and materials design.
At the same time, in the last few years, machine-learning (ML) potentials have advanced to the point where they can predict the energy landscape at a fraction of the cost of first-principles calculations, yet with comparable accuracy~\cite{Batzner2022,  Vandermause2022, Musaelian2023}.
Therefore, ML potentials can reliably replace \emph{ab-initio} calculations within the \emph{interpolative} region of configuration space represented by their training set (without exceeding the level of theory they are trained on).
Active-learning strategies make this replacement efficient and reliable by detecting when new first-principles data is needed to improve the ML potential in low-fidelity regions when \emph{extrapolation} is occurring~\cite{Jinnouchi2019, Bernstein2019, Vandermause2020, Lysogorskiy2023, Yang2024, Belli2025, Solovykh2025}.
Nowadays, the majority of active-learning strategies make use of molecular dynamics (MD), which requires the integration of the equations of motion to generate the atomic configurations.
This has two main drawbacks: (i) new structures are generated sequentially and are highly correlated, so that consecutive snapshots carry little additional information for training, and their generation cannot be parallelized, and (ii) the dynamics may become unphysical if the ML potential is not sufficiently accurate.
In contrast, within SSCHA, configurations can be stochastically drawn directly from the nuclear density matrix~\cite{Errea2014} at a cost that is negligible compared to their labelling, and without integrating any equation of motion.
Notably, the generated configurations are statistically independent.
Together with the fact that an arbitrary number of them can be generated at once, and therefore labelled in parallel, this offers a natural route to novel active-learning approaches.
In fact, an approximately 96\% reduction in first-principles workload has already been reported for \ce{PdCuH2}~\cite{Belli2025} at fixed lattice geometry and temperature by pairing active learning moment tensor potentials with SSCHA.

In this work, we present an on-the-fly active-learning scheme that combines the SSCHA method~\cite{Monacelli2021a} with Bayesian ML potentials~\cite{Vandermause2020, Vandermause2022} to greatly accelerate finite-temperature calculations with first-principles accuracy.
Since Bayesian potentials natively quantify their own predictive uncertainty, the scheme automatically identifies configurations lying outside the training domain and triggers new first-principles calculations only when needed. 
This method can be efficiently applied to a wide range of temperatures and captures challenging phase transitions.
With this approach, we showcase here two interesting energy materials with prototypical phase transitions with implications in solid-state batteries and photovoltaic research: \ce{Li2O} and \ce{CsPbI3}. 
In particular, \ce{Li2O} is the simplest lithium oxide superionic conductor, present in amorphous solid-state electrolytes~\cite{Zhang2023} and possessing a large thermal expansion, and \ce{CsPbI3} is a prototypical inorganic metal halide perovskite, extensively studied due to its exceptional optoelectronic properties for solar cell applications~\cite{Wang2019, Yoon2021, Ke2021}.
We first demonstrate how this framework can lower the SSCHA first-principles workload by 98.4 -- 99.8\% while reproducing with high fidelity the \emph{ab-initio} reference thermodynamic results.
Second, we apply the method to predict (i) the linear thermal expansion coefficient of \ce{Li2O}, and (ii) the experimental phase diagram of \ce{CsPbI3}, where the spontaneous competition between the black ($\alpha$) and yellow ($\delta$) phases is crucial in determining the light absorption properties of the material.

% ============= RESULTS
\section{Results}
% --- ACTIVE LEARNING WORKFLOW
\begin{figure*}[!]
    \centering
    \includegraphics[width=1.0\textwidth]{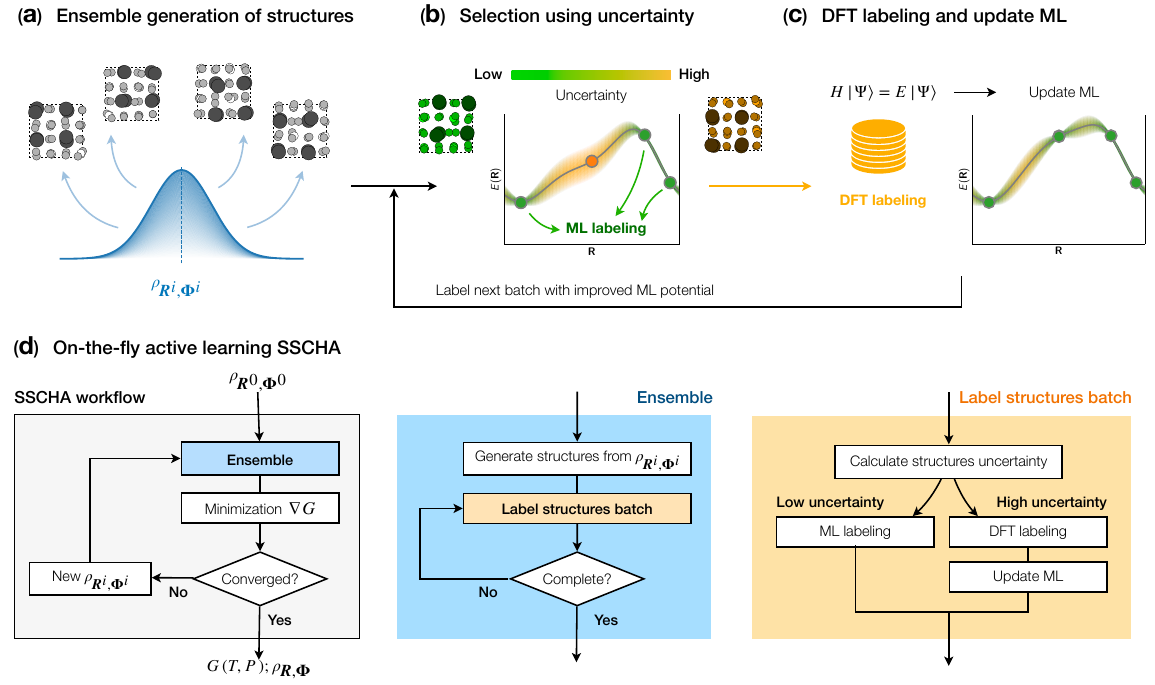}
    \caption[SSCHA on-the-fly active-learning scheme]{
        \textbf{Schematic overview of the presented  on-the-fly active learning scheme.}
        Top panels, left to right, the proposed active learning scheme involves:
        \textbf{(a)} the generation of an ensemble of structures from the nuclear Gaussian density matrix $\rho_{\mathbf{R}^i,\bm{\Phi}^i}$ at iteration $i$;
        \textbf{(b)} the selection of structures based on ML uncertainty, using a sparse Gaussian process force field;
        \textbf{(c)} DFT labeling, i.e., performing ground-state calculation to assign total energy $E$, forces, and stress of the selected structures, and use these to update the ML potential.
        Panel \textbf{(d)} schematically represents the on-the-fly active learning SSCHA with its nested processes.
        The left gray panel shows the SSCHA minimization workflow, which takes as input an initial guess density matrix $\rho_{\mathbf{R}^0,\bm{\Phi}^0}$, and self-consistently minimizes the (Gibbs) free-energy $G$ at a specific temperature $T$ and pressure $P$, providing the converged final density matrix $\rho_{\mathbf{R},\bm{\Phi}}$.
        In blue, the generation of the ensemble of structures, which are iteratively labeled by splitting these into small batches.
        The labeling of each batch (orange panel) consists of using the ML potential for structures where the model is confident of its predictions, and single-point DFT calculations otherwise, which are used to improve the ML potential for labeling the next batch.
    }
    \label{fig:workflow}
\end{figure*}
% -------
% =========================================================================================== %
\subsection{On-the-fly SSCHA active learning}
% =========================================================================================== %
%
The SSCHA~\cite{Errea2013, Monacelli2018, Monacelli2021a} is a method to compute the quantum free energy of the atoms in a molecule or a solid as the minimum of the functional $F_{T}[\rho]$ of a trial nuclear density matrix $\rho$, with the assumption that $\rho$ can be represented by a multivariate Gaussian.
The latter is parametrized in terms of nuclei centroid positions $\mathbf{R}$ and effective interatomic force constants (IFC) $\bm{\Phi}$:
\begin{equation}
    {\rho}_{\mathbf{R},\bm{\Phi}}(\mathbf{r})
    =
    \sqrt{\mathrm{det} \left (\frac{\bm{\Psi}^{-1}}{2\pi} \right )}
    e^{
    %\left \{
        - \frac{1}{2}
        \sum_{ab} 
        (r_a-R_a) 
        \Psi^{-1}_{ab}
        (r_b -R_b)
    %\right \}
    }
    ~,
    \label{eq:solids:gaussianrho}
\end{equation}
where
\begin{equation}
    \Psi_{ab}
    =
    \frac{1}{\sqrt{M_a M_b}}
    \sum_{\mu}
    \frac{\hbar (2n_{\mu} + 1)}{2\omega_{\mu}}
    e^{a}_{\mu}e^{b}_{\mu}
    ~;
    \label{eq:solids:psi}
\end{equation}
here, $\omega_{\mu}$ and $\mathbf{e}_{\mu}$ are the frequencies and eigenvectors of the mass-rescaled effective IFC $\bm{\Phi}$, $n_{\mu}$ the Bose-Einstein occupation, and $\mathbf{r}$ are the positions of the nuclei.
Minimizing $F$ requires the calculation of its gradient with respect to $\mathbf{R}$  and $\bm{\Phi}$. In SSCHA~\cite{Errea2013, Errea2014, Monacelli2018, Monacelli2021a}, the required quantities are calculated as stochastic integrals, which are evaluated by generating an \emph{ensemble} of structures (Fig.~\ref{fig:workflow}(d)) and labeling them (i.e., computing total energy, forces, and stress tensors).
The atomic coordinates of these structures are extracted from the trial Gaussian density $\rho_{\mathbf{R},\bm{\Phi}}$ (Figs.~\ref{fig:workflow}(a) and (d)) by generating for each structure a vector of random numbers $y_{\mu}$ and employing it as follows~\cite{Errea2014}:
\begin{equation}
    r_a = R_a+\sum_{\mu}\frac{1}{\sqrt{M_a}}
     y_{\mu} \sqrt{(2n_{\mu} + 1)\frac{\hbar}{2\omega_{\mu}}} e_{\mu}^{a}
     ~.
    \label{eq:generation}
\end{equation}
As the labeling represents the most computationally demanding step of the SSCHA method, the main objective is to substitute the first-principles labeling, typically from density-functional theory (DFT) calculations, with a computationally inexpensive ML potential.

We schematically outline in Fig.~\ref{fig:workflow} the proposed on-the-fly active-learning workflow and its principal components.
To understand whether the ML model is reliable to label a specific structure so as to replace the DFT calculation, we employ sparse Gaussian process (SGP) force fields (Fig.~\ref{fig:workflow} (b)), as implemented in the \texttt{FLARE} code~\cite{Vandermause2020,Vandermause2022}.
SGP models have the ability to estimate the uncertainty of their own predictions, which, alongside a user-defined threshold, gives a natural and automated way to know whether the model can be reliably used~\cite{Vandermause2020,Vandermause2022}.
When the uncertainty exceeds the threshold for an atomic configuration, the latter is \emph{selected} (Fig.~\ref{fig:workflow}(b)) and a ground-state DFT calculation is called for the labeling and then used to refine the SGP force field (Fig.~\ref{fig:workflow}(c)).

The SGP model allows for selecting the uncertain structures over the whole ensemble, in order to refine the model.
However, this approach is not the most efficient: re-training the model using only a few new structures could be enough to obtain the required accuracy.
This is true when the structures in the ensemble share reasonably similar atomic environments, as is likely the case for structures generated using Eq.~\ref{eq:generation}.

To trade off between the number of unnecessary DFT calculations, the number of re-training calculations, and the time-to-solution, we divide the ensemble into $N$ \emph{batches}, each containing $N_c / N$ configurations, where $N_c$ is the total number of structures in each ensemble.
For each batch, we evaluate the uncertainty of the ML and select the uncertain structures, label them using DFT, and then update the SGP model with these labels (blue and orange panels in Fig.~\ref{fig:workflow}(d)). This process is done for each batch, using each time the new updated model.
We note that the SSCHA minimization usually requires a large number of configurations to obtain accurate free energies~\cite{Monacelli2021a}. 
This can be readily obtained with our strategy, as each round of DFT calls is done on the batches, and therefore the on-the-fly process is limited to a smaller fraction of configurations only, at variance with other approaches~\cite{Belli2025}.
Importantly, we stress that this procedure allows us to improve the SGP model from batch to batch, as shown in Figs.~\ref{fig:workflow}(b) and (c), thus minimizing the probability of finding uncertain structures in the subsequent batches, thereby lowering the number of DFT calculations.
The number of batches $N$ is a user-defined input; in our numerical experiments, we found that having about 10 structures in each batch provides a good trade-off between wall time and number of unnecessary DFT calls.

As one is often interested in performing several SSCHA calculations at different temperatures -- for instance, when computing the thermal expansion of a material or the phase transition temperature between two competing phases -- we also adopt a strategy to accelerate the SSCHA minimization and reduce the total number of structures to be labelled (see Appendix~\ref{sec:guessstrategy}). 
The latter strategy is fundamental to maintain good inference speed for the SGP model, whose performance depends on the number of training labels~\cite{Bartok2010, Vandermause2022}.
%
% --- EFFICIENCY AND ACCURACY
\begin{figure*}
    \centering
    \includegraphics[width=1.00\textwidth]{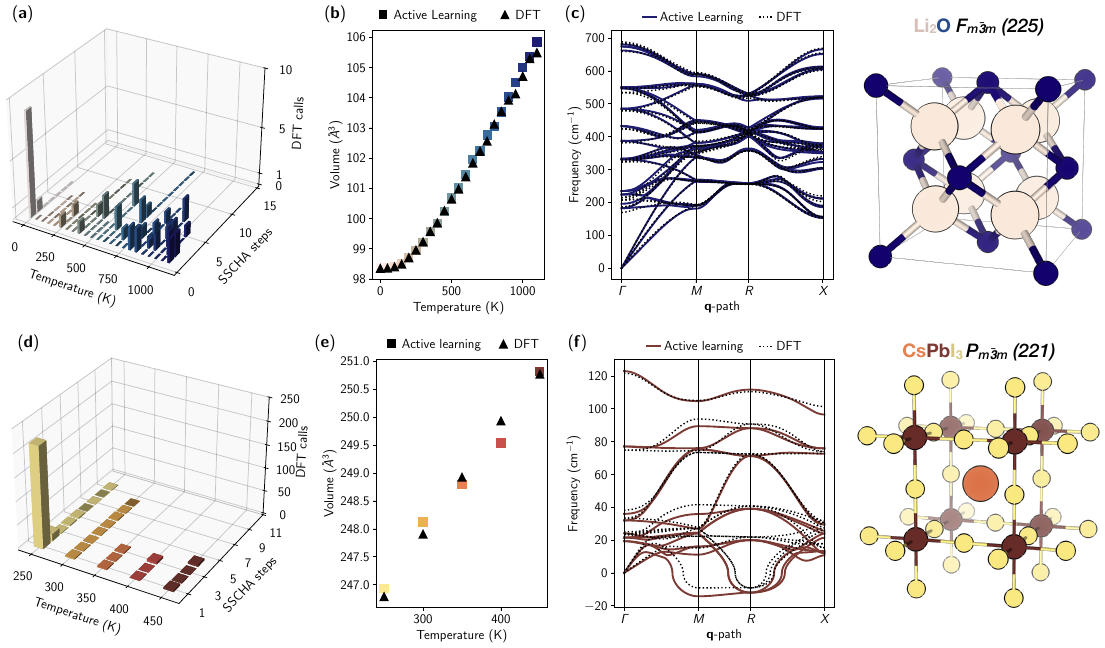}
    \caption{
        \textbf{Thermodynamic and dynamical properties from the proposed active learning approach.}
        Results of the present active-learning framework applied to \textbf{(a -- c)} \ce{Li2O} and to \textbf{(d -- f)} $\alpha$--\ce{CsPbI3} to calculate the equilibrium volume as a function of temperature, and compared to reference SSCHA calculations carried out entirely with DFT (SSCHA-DFT).
        \textbf{(a,d)} Number of DFT labeling calls during the on-the-fly training as a function of temperature and number of SSCHA minimization steps. 
        Each iteration required the labeling of 200 and 400 configurations, for \ce{Li2O} and $\alpha$--\ce{CsPbI3} respectively.
        \textbf{(b,e)} Finite-temperature SSCHA equilibrium volume as obtained from the on-the-fly active learning procedure and compared to the reference SSCHA-DFT calculations.
        \textbf{(c,f)} Hessian of the positional free energy $F(\mathbf{R})$ (see Appendix~\ref{sec:hessian}) dispersion at 450~K, as obtained from the presented on-the-fly active learning scheme and compared to the reference DFT calculations.
    }
    \label{fig:alvsdft}
\end{figure*}
% ---
% =========================================================================================== %
% =========================================================================================== %
\subsection{Efficient and accurate thermodynamic properties}
\label{sec:solids:application}
% =========================================================================================== %
% =========================================================================================== %
First, we use the method presented to calculate the thermodynamic properties of \ce{Li2O} and $\alpha$--\ce{CsPbI3}, demonstrating the efficiency of the proposed approach.
All the calculations reported in this section are performed in supercells commensurate with the DFPT $\mathbf{q}$-point meshes, containing 96 atoms ($2\times2\times2$ the conventional unit cell) for \ce{Li2O} and 40 atoms ($2\times2\times2$) for $\alpha$--\ce{CsPbI3} (see Appendix~\ref{sec:compdet}).
In Figures~\ref{fig:alvsdft}(a) for Li$_2$O and (d) for $\alpha$-CsPbI$_3$, the number of DFT calculations is shown as a function of the SSCHA minimization step~\cite{Note1} and temperature.
We first notice that the majority of the DFT labeling happens at the very first SSCHA step and at the initial (lowest) temperature. The number of DFT labeling calls generally decreases with increasing SSCHA steps and temperatures. Notably, DFT calls are overall much fewer than the total number of configurations generated for each SSCHA step, which are 200 and 400 for \ce{Li2O} and $\alpha$--\ce{CsPbI3}, respectively.
In particular, at the end of the active learning, only 44 (\ce{Li2O}) and 256 ($\alpha$-\ce{CsPbI3}) DFT calculations were required to fit a reliable ML potential.
We stress that these numbers refer to the complete active-learning workflow, i.e., to the entire temperature ramp: the SSCHA calculations are performed sequentially, from the lowest to the highest temperature, each one starting from the previously converged solution (see Appendix~\ref{sec:guessstrategy}), and the DFT calls resolved per temperature and SSCHA step in Figs.~\ref{fig:alvsdft}(a) and (d) sum up to the quoted totals.
For each system, a single SGP model is trained on the fly and transferred across all temperatures and volumes, without ever retraining it from scratch; thermal expansion is naturally accounted for during training, since the variable-cell SSCHA relaxations expose the model to the atomic environments of the expanded lattices.
Importantly, our scheme is more than twice as efficient as previous on-the-fly MD simulation of $\alpha$-\ce{CsPbI3}~\cite{Jinnouchi2019}, where 520 DFT calculations were needed to perform the simulation at 700~K only.
In passing, we note that the generation of the structures with the proposed approach is carried out inexpensively without resorting to any dynamics integration, unlike MD-based active-learning schemes. 
To better understand the efficiency, our reference SSCHA-DFT calculations in Fig.~\ref{fig:alvsdft}(b) and (e) needed 21'347 and 16'000 DFT labels, respectively, for \ce{Li2O} and $\alpha$-\ce{CsPbI3}.
This proves that our scheme allows a reduction of 98.4\%--99.8\%  in first-principles workload.
Moreover, it is worth noting that the 44 DFT calculations required for the active-learning of \ce{Li2O} are comparable in number to those needed for a quasi-harmonic approximation calculation (see Supplementary Discussion), which, however, does not capture full anharmonic effects.

All these considerations indicate that SSCHA, coupled with the active-learning scheme, enables a targeted exploration of the potential energy surface.
This allows not only a highly efficient generation of the ML potential, but also proves that the relevant physics is contained in only a few configurations of the phase space.
A focused approach, as the one proposed here, is therefore much more effective if one targets specific physical properties such as thermodynamic free energies, compared to methods that aim at exploring a much larger phase-space~\cite{Liu2025}.
Notably, the approach shows that an increase in the number of DFT calls mainly happens toward phase transitions, as can be noticed for Li$_2$O in Fig.~\ref{fig:alvsdft}(a) at high temperatures close to the experimental superionic transition at around 1200~K~\cite{Gupta2012}, and it is therefore sensitive to the relevant physics of the material.

In Fig.~\ref{fig:alvsdft}(b) and (e), we show the accuracy of the active learning and the underlying ML potentials by comparing the temperature-dependent equilibrium volume with our computed reference SSCHA-DFT calculations.
The comparison shows that the ML potentials reproduce with high fidelity the reference SSCHA-DFT results -- the small discrepancies may be due to the remaining stochastic noise of the SSCHA method, and one should not infer that SSCHA-DFT is necessarily more accurate.
Indeed, the volumes differ by no more than $\sim 0.1$~\AA$^3$, a discrepancy that likely reflects the residual statistical uncertainty in the SCHA stress tensor, whose convergence to within this tight threshold requires a large number of configurations~\cite{Monacelli2018, Monacelli2021a}.
From Fig.~\ref{fig:alvsdft}(b), we obtain the linear thermal expansion coefficient $\alpha_{T_0}$ at temperature $T_0$ for Li$_2$O, which we compute as follows:
\begin{equation}
    \alpha_{T_0} = \frac{1}{a_{T_0}} \left . \frac{\partial a}{\partial T} \right |_{T=T_0}
    ~,
    \label{eq:linearcoeff}
\end{equation}
where $a$ is the lattice parameter of the unit cell.
To obtain a smooth function for $a$ as a function of the temperature, we fit the lattice parameters obtained from the active learning using a fourth-order polynomial (see Supplementary Figure 1).
Employing the fit in Eq.~\ref{eq:linearcoeff}, $\alpha_{T_0}$ results in $\SI{2.8e-5} {\per\kelvin }$ at $T_0 = 1000$~K, which is in excellent agreement with the experimental values of about $\SI{3e-5} {\per\kelvin }$ \cite{Yao2013,Kurasawa1982}.

Figs.~\ref{fig:alvsdft}(c) and (f) report the phonon dispersions of Li$_2$O and $\alpha$-CsPbI$_3$ at 450~K.
These phonon band structures correspond to the Hessian matrix dispersion, i.e., the second-order derivative of the positional free energy corresponding to the observable vibrational frequencies (see Appendix~\ref{sec:hessian}), which is typically more sensitive to the accuracy of the forces than the auxiliary harmonic dispersion.
From these plots, we notice that Li$_2$O is dynamically stable in the \textit{Fm$\overline{3}$m} phase, as all its phonon frequencies are positive (the corresponding crystal structure is represented on the right-hand side of Fig.~\ref{fig:alvsdft}(c)). 
On the contrary, \ce{CsPbI3} shows imaginary phonons (negative in the plot) at the $R$ and $M$ points of the first Brillouin zone, signaling the instability of the material in the \textit{Pm$\overline{3}$m} phase at $T = 450$~K (the related crystal structure is depicted in the right-hand side of Fig.~\ref{fig:alvsdft}(f)). 
These instabilities are due to the octahedral-tilting modes of the cubic perovskite lattice, and they drive the displacive sequence towards the tilted black variants, i.e. the tetragonal $\beta$ ($P4/mbm$) and orthorhombic $\gamma$ ($Pnma$) phases.
We stress that they are unrelated to the $\delta \rightarrow \alpha$ transition discussed in the next section, which is a reconstructive, first-order transition that cannot be signalled by soft modes of the cubic lattice~\cite{Monacelli2023}.
For this reason, later we determine it from the Gibbs free-energy balance of the two phases.
In Figs.~\ref{fig:alvsdft}(c) and (f), we also compare the active-learned and reference dispersions for both materials studied, showing excellent agreement with one another, proving that our method can accurately capture very sensitive dynamics of the crystals.
We remark that applying the active learning to the $\alpha$ phase in a temperature range where it is dynamically unstable is motivated by three considerations.
First, the SSCHA formalism allows free energies to be computed also for dynamically unstable phases, and our end goal -- addressed in the next section -- is the $\delta \rightarrow \alpha$ transition, which requires the free energy of the $\alpha$ phase rather than that of the tilted black variants.
Second, it has been shown in Ref.~\citenum{Monacelli2023}, for the analogous \ce{CsSnI3} material, that the cubic $\alpha$ phase is a good proxy for the other perovskite phases, $\beta$ and $\gamma$, since the dynamics of the relevant octahedral-tilting modes is already encoded in its anharmonic spectral function.
Third, the softness of these tilting modes makes the free-energy Hessian particularly sensitive to the accuracy of the ML potential, since small errors in the potential translate into large relative changes of its near-vanishing eigenvalues; this comparison therefore provides a stringent benchmark of the active-learning scheme against the SSCHA-DFT reference.

Importantly, all the results shown in Fig.~\ref{fig:alvsdft} for the active learning are taken as given directly by the on-the-fly SSCHA simulations, without further retraining the ML potentials or repeating the SSCHA minimization with an updated model.
%

% --- APPLICATION
\begin{figure}
    \centering
    \includegraphics[width=.45\textwidth]{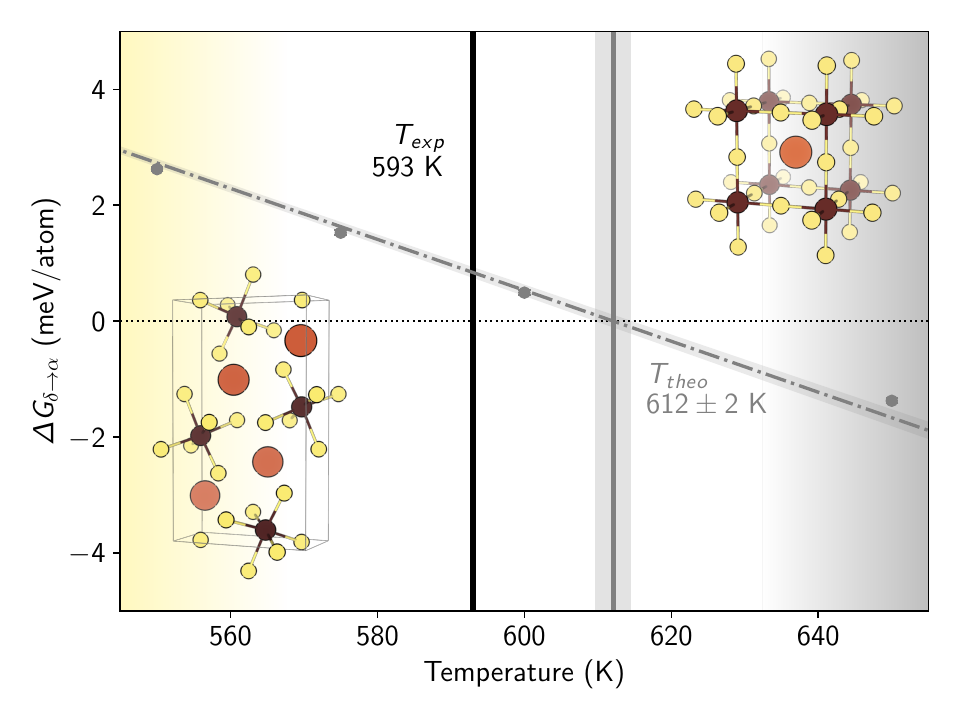}
    \caption{
        \textbf{$\delta \rightarrow \alpha$ phase transition of \ce{CsPbI3}.}
        SSCHA Gibbs free energy difference between $\delta$ and $\alpha$ phases of \ce{CsPbI3} as a function of temperature, represented by the gray bullets with their stochastic error bars. The gray dash-dotted line is the weighted linear fit used to determine the transition temperature, shown with its $1\sigma$ confidence band (see Supplementary Discussion). The crossing $\Delta G_{\delta\rightarrow\alpha}=0$ indicates the predicted phase transition temperature $T_{\mathrm{theo}} = 612 \pm 2$~K, highlighted by the vertical gray line and its $\pm 1\sigma$ shaded band, and it is compared with the experimental temperature $T_{\mathrm{exp}} \simeq 320 {}^\circ C = 593$~K from Ref.~\citenum{Ke2021}. The crystal structures of the $\delta$ (bottom left) and $\alpha$ (top right) phases are also depicted.
    }
    \label{fig:phasetransitions}
\end{figure}
% ---

% =========================================================================================== %
\subsection{$\delta - \alpha$ phase transition in \ce{CsPbI3}}
% =========================================================================================== %
%
The black perovskite phase of \ce{CsPbI3} has received significant attention in the 
context of photovoltaic technology due to its favorable band gap~\cite{Wang2019, Yoon2021, Ke2021}.
However, this phase, also referred to as $\alpha$, is dynamically unstable at room temperature, and spontaneously degrades to the non-perovskite $\delta$ phase.
As the $\alpha$ phase becomes thermodynamically accessible only above 593~K~\cite{Marronnier2018,Ke2021}, finding ways to stabilize it at room temperature is extremely important.
Here we focus on predicting the $\delta \rightarrow \alpha$ phase-transition temperature using the SSCHA method, which constitutes an important step toward engineering strategies that shift the thermodynamic balance between competing polymorphs.

The phase-transition temperature can be obtained from the Gibbs free energy difference between the $\delta$ and $\alpha$ phases, $\Delta G_{\delta\rightarrow\alpha} = G_{\alpha} -G_{\delta}$, and corresponds to the temperature $T$ for which  $\Delta G_{\delta\rightarrow\alpha}(T) = 0$.
To this end, SSCHA is employed to find the Gibbs free energy~\cite{Monacelli2018} for both phases, where $G = F+\Omega P$, $P$ is the hydrostatic pressure, and $\Omega$ is the unit cell volume.
We minimize $G$ using variable-cell SSCHA optimization, where the lattice vectors are also relaxed while keeping the pressure fixed (see Appendix~\ref{sec:sschainfo}).
In this context, the calculations are carried out at zero pressure ($P=0$), representative of ambient conditions.

In the previous section (see Fig. \ref{fig:alvsdft}(d)-(f)), we trained an ML potential that accurately reproduces the SSCHA-DFT reference results in the temperature range 250 -- 450~K for the $\alpha$ phase.
As we are not sure about the accuracy of this ML potential in describing the $\delta$ phase, we apply our active-learning scheme in the same temperature range, 250 -- 450~K, to the $\delta$ phase.
As the SGP model prediction performance decreases with the number of training data, we do not make use of the previous ML potential for the $\alpha$ phase as a starting point for the active learning of the $\delta$ phase.
By training from scratch a new SGP model during the active-learning SSCHA calculation for the $\delta$ phase, the simulation still only required 50 DFT calculations.
At this point, we have 256 and 50 total energy calculations for the $\alpha$ and $\delta$ phases, respectively, making up a dataset composed of 306 DFT labels.
To create a fast SGP model that can work for both phases, we reduce the number of configurations in the dataset in such a way as to select configurations having the maximum force component acting on atoms below or equal to 10~eV/\AA.
The latter value is chosen after noticing that the force components in the atomic configurations extracted during the converged SSCHA calculations never exceed this threshold.
This reduces the training dataset to 172 structures, and it is used to train a single SGP model, which is then used for both phases, and across all temperatures and volumes, in the Gibbs free-energy calculations that follow.
The model thus trained reproduces the DFT potential energy surface with high fidelity, as verified by inspecting parity plots of total energy and forces (see Supplementary Figure 2) and their mean absolute errors (see Supplementary Table I), which are comparable to previous studies on quantum paraelectric materials~\cite{Ranalli2023}.

In Fig.~\ref{fig:phasetransitions}, we show the final Gibbs free energy difference $\Delta G_{\delta \rightarrow \alpha}$ as a function of temperature computed with this ML potential and SSCHA (see Appendix~\ref{sec:sschainfo}).
Notably, the free energies entering $\Delta G_{\delta \rightarrow \alpha}$ are computed in the 450--650~K range, above the 250--450~K window in which the potential was trained.
The reasonable prediction of the transition temperature therefore also probes the extrapolative power of the ML potential towards higher temperatures, although we note that a strict benchmark would be to compare the transition temperature against a reference SSCHA-DFT calculation, which, however, is prohibitively expensive for such large supercells.
The Gibbs free energies of the two phases are computed in supercells containing up to 2560 atoms ($8\times8\times8$ for $\alpha$ and $4\times8\times4$ for $\delta$), and the transition temperature is carefully converged with respect to both the supercell size and the number of configurations in the ensemble (see Supplementary Figure~4, Supplementary Table II, and Supplementary Discussion).
To estimate the transition temperature $T_{\mathrm{theo}}$, we find the zero of $\Delta G_{\delta \rightarrow \alpha}(T)$ with a linear fit through all the computed temperatures, where each point is weighted by the inverse square of its stochastic error, and the uncertainty on $T_{\mathrm{theo}}$ is propagated from the fit (see Supplementary Discussion).
This temperature is highlighted by the vertical gray line in Fig.~\ref{fig:phasetransitions}, and corresponds to $T_{\mathrm{theo}} = 612 \pm 2$~K, which is within 20~K of the experimental reference value $T_{\mathrm{exp}} = 593$~K~\cite{Ke2021}.
This agreement indicates that the SSCHA, combined with an accurate model of the potential energy surface, provides a predictive estimate of the solid-solid transition temperature.
%

% ============================= DISCUSSION
\section{Discussion}
The discovery and engineering of materials remain a central challenge for sustainable technological development and a green transition. 
As increasingly sophisticated tools for materials discovery emerge, including generative artificial intelligence~\cite{Zeni2025}, the present framework provides a key step towards reliable and efficient finite-temperature atomistic modeling.
This framework is implemented as an on-the-fly active-learning workflow within the SSCHA method. 
The combination of stochastic sampling and predictive uncertainties reduces the first-principles workload by two to three orders of magnitude, while accounting for the finite-temperature relaxation of all the structural degrees of freedom, i.e., atomic positions and lattice vectors.
Moreover, it allows configurations to be labelled in parallel and remain confined to the relevant region of phase space even when the ML potential is still inaccurate, at variance with sequential sampling, such as molecular dynamics, where an imperfect potential can drive the trajectory towards ``unphysical'' regions. 
We showcased the proposed approach in accurately predicting the extreme thermal expansion of \ce{Li2O}, and the challenging phase-transition temperature governing the exceptional optoelectronic properties of \ce{CsPbI3}, namely the transition from the $\delta$--yellow to the $\alpha$--black phase.
For \ce{Li2O}, we obtain a linear thermal expansion coefficient of $2.8\times10^{-5}$~K$^{-1}$ at 1000~K, in line with the experimental value of about $3\times10^{-5}$~K$^{-1}$; for \ce{CsPbI3}, a transition temperature of $612$~K, less than 20~K from the experimental value of 593~K.

Taken together, this work establishes two main points.
First, we demonstrate that the essential physics governing crystal thermodynamics is encoded in a restricted region of phase space, which contains sufficient information to train an accurate ML potential. This is established by accurately reproducing finite-temperature vibrational properties using ML potentials trained on a remarkably small set of physically meaningful configurations.
Second, we provide a framework to efficiently generate and select these meaningful configurations: stochastic sampling of the nuclear density matrix generates physically meaningful configurations, while a Bayesian ML potential selects those that provide nonredundant information. 

These findings broaden the scope of finite-temperature simulations and provide a practical route to training and fine-tuning foundation ML potentials.
First and foremost, the demonstrated efficiency brings within reach levels of electronic-structure theory, e.g., the random phase approximation~\cite{Ren2012} or quantum Monte Carlo~\cite{Foulkes2001}, that would otherwise be prohibitive to employ at finite temperature, and which are crucial for describing phenomena such as quantum paraelectricity in correlated materials~\cite{Verdi2023} and matter under extreme pressure~\cite{Monacelli2020, Monacelli2023a}.
In this sense, the framework opens the way not only to cheaper predictions, but to physics that is presently out of reach.
A second aspect is connected to foundation neural-network potentials, such as MACE-MP~\cite{Batatia2025a}, PET-MAD~\cite{Mazitov2025}, or NequIP/Allegro~\cite{Kavanagh2026}, that promise increasing transferability.
In this respect, we see the present framework to be valuable in two scenarios.
On the one hand, it provides a recipe for the inexpensive generation of large, but compact, datasets for training foundation ML potentials for accurate thermodynamic and vibrational properties.
On the other hand, it can be combined with the fine-tuning of foundation ML potentials for targeted applications that require expensive first-principles methods. 
In either case, the present scheme resolves the bottleneck of selecting which configurations should be labelled for evaluating anharmonic lattice dynamics.
In fact, while more expressive neural-network architectures~\cite{Batatia2022,Batzner2022,Musaelian2023} are expected to further reduce the number of labels required to reach a target accuracy, they do not, by themselves, indicate which configurations should be computed, nor do they generally provide the native error estimate that drives the on-the-fly selection, although notable progress is being made in this direction~\cite{Grasselli2025}.
Instead, when trained or fine-tuned on the compact datasets generated by the proposed approach, such models are expected to yield potentials even more accurate than the sparse Gaussian process employed in this work, as already demonstrated in other contexts~\cite{Vandermause2022,Xie2023,Perego2024,Falletta2025,Rivano2026}.

The combination of efficiency and predictive reliability has profound implications for the accurate engineering of energy materials, making the proposed approach appealing for a wide range of technologically relevant applications, including batteries, photovoltaics, thermoelectrics, and superconductors.
More broadly, the present scheme can be extended to accelerate the calculation of other response properties, including infrared and Raman spectra~\cite{Monacelli2021,Miotto2024,Bastonero2024}, as well as out-of-equilibrium quantum dynamics~\cite{Libbi2025,Libbi2025a} within the time-dependent SCHA formalism~\cite{Monacelli2021}. 
In this way, it could provide a unique and self-contained framework for accelerated materials characterization.

The implementation is open source and available in the \texttt{python-sscha}~\cite{Monacelli2021a} code, and relies on the \textsc{FLARE} code~\cite{Vandermause2020, Vandermause2022} for the Bayesian potential, and on the AiiDA infrastructure~\cite{Pizzi2016,Huber2020,Uhrin2021} for automated and reproducible calculations, following the FAIR principles~\cite{Wilkinson2016}.
We hope that these features, along with the demonstrated accuracy and efficiency of the method, will help further support computational materials research under realistic conditions, even more so when only limited access to supercomputing facilities is available.

% ============================= Data availability
\section{Data availability}
The data used to produce the results of this work are available on the Materials Cloud Archive~\cite{Bastonero2026a,Talirz2020}.

% ============================= Code availability
\section{Code availability}
All the code developed and presented here is open source and made available on GitHub as part of the official \texttt{python-sscha} distribution (\href{https://github.com/SSCHAcode/python-sscha}{https://github.com/\-SSCHAcode/\-python-sscha}), and supported from version 1.7. 
The AiiDA infrastructure~\cite{Pizzi2016, Huber2020, Uhrin2021} (\href{https://github.com/aiidateam/aiida-core}{https://github.com/\-aiidateam/\-aiida-core}), the \texttt{aiida-quantumespresso} plugin~\cite{Huber2021,Huber2026} (\href{https://github.com/aiidateam/aiida-quantumespresso}{https://github.com/\-aiidateam/\-aiida-quantumespresso}), the \texttt{aiida-quantumespresso-ph} plugin (\href{https://github.com/aiidateam/aiida-quantumespresso-ph}{https://\-github\-.com/\-aiidateam/\-aiida-\-quantumespresso-ph}), the Quantum ESPRESSO code suite~\cite{Giannozzi2009, Giannozzi2017, Giannozzi2020} (\href{https://gitlab.com/QEF/q-e}{https://\-gitlab.com/\-QEF/\-q-e}), and the FLARE code~\cite{Vandermause2020, Vandermause2022} (\href{https://github.com/mir-group/flare}{https://github.com/\-mir-group/\-flare}) are all open-source software.

% ============================= ACKNOWLEDGEMENTS
\section{Acknowledgements}
L.B., C.C., and N.M. gratefully acknowledge support from the Deutsche Forschungsgemeinschaft (DFG) under Germany’s Excellence Strategy (EXC 2077, No. 390741603, University Allowance, University of Bremen) and Lucio Colombi Ciacchi, the host of the ``U Bremen Excellence Chair Program''.
G.J. gratefully acknowledges the support from the Brittany region (JALI grant, No. 5193, Université de Rennes).
We acknowledge support by the NCCR MARVEL, a National Centre of Competence in Research, funded by the Swiss National Science Foundation (Grant number 205602). 
This work was supported by a grant from the Swiss National Supercomputing Centre (CSCS) under project ID~465000416 (LUMI-G).
%

% ============================= AUTHOR CONTRIBUTION STATEMENT
\section{Author contributions statement}
We use in the following the CRediT (Contributor Roles Taxonomy) author statement.
L.B.: conceptualization, methodology, software, validation, formal analysis, data curation, writing -- original draft, visualization;
G.J.: validation, software, methodology, formal analysis;
C.C.: validation, methodology, visualization;
L.M.: conceptualization, supervision, methodology, software;
N.M.: conceptualization, supervision, project administration, funding acquisition.
All authors: writing -- review \& editing.

% ============================= ADDITIONAL INFO
\section{Competing Interests}
The authors declare no competing interests.

% ============================= METHODS
\appendix\label{Appendix}
% =========================================================================================== %
\section{Theoretical and workflows details}
% =========================================================================================== %

% =========================================================================================== %
\subsection{Optimal-guess strategy}
\label{sec:guessstrategy}
Often, one is interested in performing several SSCHA calculations at different temperatures; for instance, when computing the thermal expansion of a material or the phase transition temperature between two phases, as in the present work.
To make these calculations even more efficient, we adopt another important strategy during the active-learning simulation that exploits the capabilities of the SSCHA framework.

We imagine carrying out SSCHA calculations in a ramping temperature range, starting from an initial temperature $T^0$, until a final temperature $T^{\mathrm{max}}$, with a temperature step $\Delta T$.
If $\Delta T$ is small (e.g., $\lesssim 50$~K), we can expect that the density matrix at a generic temperature step $i$, $\tilde{\rho}(T^i)$, will be similar to the density matrix at the next step $i+1$, $\tilde{\rho}(T^{i+1} = T^i+\Delta T)$.
This means that we can use the converged centroids and effective IFC of the SSCHA calculation at $T^{i}$ as the guess for the next calculation at $T^{i+1} = T^{i} + \Delta T$, with the hope of reducing the number of iterations required to minimize the free energy, as well as reducing the probability of generating new uncertain atomic environments (in passing, we note that the density matrix guess from the previous iteration can be, at almost no cost, further improved by employing the importance-sampling technique implemented within SSCHA~\cite{Monacelli2021a}).
Importantly, similar density matrices generate similar ensemble configurations, meaning that the generated atomic environments will be highly correlated.
As a consequence, applying this ``optimal-guess'' strategy is fundamental to reduce the number of steps during the free energy minimization and to avoid the generation of out-of-distribution atomic configurations, which could ultimately increase the number of DFT calls (and even decrease the ML potential performance). 

% =========================================================================================== %
\subsection{Hessian of the positional free energy}
\label{sec:hessian}
% =========================================================================================== %
An important remark is that the \emph{auxiliary} IFC $\Phi$ resulting from the SSCHA minimization has no physical relation to the ``real'' phonons, as observed for example in neutron scattering experiments, as it represents one of the parameter to define the trial density matrix $\rho_{\mathbf{R}, \bm{\Phi}}$.
To obtain the physical phonon frequencies, one computes the Hessian of the \emph{positional free energy}, $F({\mathbf{R}})$, i.e., the second-order derivative of $F$ with respect to \textbf{R}:
\begin{equation}
H_{ab}(R) = \frac{1}{\sqrt{ M_a M_b}} \cdot \frac{\partial^2 F}{\partial R_a \, \partial R_b}.
\end{equation}
In fact, this second-order derivative of $F(\textbf{R})$ can be considered as a natural generalization of the harmonic dynamical matrix, now including thermal and quantum nuclear effects.
The eigenvalues of this Hessian matrix tend towards zero as the system approaches an instability. 
This property enables the study of second-order phase transitions by inspecting the evolution of eigenvalues with temperature to determine the critical transition temperature.
In practical terms, the Hessian is solved using an analytical formulation, which reads:
\begin{equation}
    \frac{\partial^2 F}{\partial R_a \, \partial R_b}
    = 
    \Phi_{ab}
    + \sum_{cdef} \overset{(3)}\Phi_{acd}
    \left[ \mathbb{1} - \Lambda\, \overset{(4)}\Phi \right]^{-1}_{cdef}
    \overset{(3)}\Phi_{efb},
\end{equation}
where $\overset{(n)}{\Phi}$ in the $n$-order force constant, from which the dynamical matrix can be calculated as $(\overset{(n)}\Phi)_{ab..n}/
\sqrt{M_aM_b...M_n}$, and $\Lambda$ depends on SSCHA eigenvalues and eigenvectors. 
We refer the reader to Ref.~\citenum{Monacelli2021a} and Ref.~\citenum{Bianco2017} for more details.
In our work, we keep up to the 3rd-order term for the calculation of the Li$_2$O Hessian, as the 4th-order is notoriously very computationally demanding and not relevant in this specific case. 
We include the 4th-order term in the study of CsPbI$_3$, which has been shown to be important for this type of structures~\cite{Monacelli2023}.
%

% =========================================================================================== %
\section{Computational details}
\label{sec:compdet}
% =========================================================================================== %

% =========================================================================================== %
\subsection{Density-functional theory (DFT)}
% =========================================================================================== %
%
Calculations are performed using the \textsc{Quantum \-ESPRESSO} distribution v7.2~\cite{Giannozzi2009, Giannozzi2017, Giannozzi2020}. 
We use the PBEsol~\cite{Perdew2008} exchange-correlation functional approximation, and employ pseudopotentials 
from the SSSP  PBEsol  precision library (version 1.3)~\cite{Prandini2018, 
Garrity2014, Schlipf2015, Willand2013, Corso2014, Topsakal2014, Setten2018} for \ce{Li2O}, and
optimized norm-conserving Vanderbilt pseudopotentials~\cite{Hamann2013} from the PesudoDojo 
library~\cite{Setten2018} for \ce{CsPbI3}.
The kinetic-energy wavefunction and charge-density cutoffs are set to the recommended SSSP  precision values for \ce{Li2O} (75~Ry and 600~Ry, respectively), and to 70~Ry and 280~Ry for \ce{CsPbI3} phases.
For the latter, a Gaussian smearing function is also used for numerical stability with a spread of $0.03$~Ry.
The starting crystal structures are optimized using the Broyden-Fletcher-Goldfarb-Shanno (BFGS) algorithm, with a convergence threshold for the total energy of $5.0\times 10^{-5}\,$Ry, for forces of $10^{-4}\,$Ry/Bohr, and for pressure of $0.01\,$kbar, and using $4 \times 4 \times 4$ $\Gamma$-centered, $8 \times 8 \times 8$ shifted, and $4 \times 8 \times 4$ shifted, Monkhorst-Pack $\mathbf{k}$-point grids to sample the Brillouin zone for \ce{Li2O}, $\alpha$-\ce{CsPbI3}, and $\delta$-\ce{CsPbI3}, respectively.
Density-functional perturbation theory (DFPT) calculations, used to obtain the starting dynamical 
matrices, dielectric and Born effective charge tensors, were performed using the \textsc{PHonon} 
code~\cite{Baroni2001} of \textsc{Quantum ESPRESSO} using a $\mathbf{q}$-point mesh of $2 \times 2 \times 2$, $2 \times 2 \times 2$, and $1 \times 2 \times 1$ respectively for \ce{Li2O}, $\alpha$-\ce{CsPbI3}, and $\delta$-\ce{CsPbI3}.
For the SSCHA calculations reported in Fig.~\ref{fig:alvsdft}, we employed supercells commensurate with the DFPT $\mathbf{q}$-meshes, rescaling the $\mathbf{k}$-points for the supercell DFT calculations according to the respective unit-cell DFT calculations.
All calculations are submitted and automatically managed by the 
\texttt{aiida-quantumespresso}~\cite{Huber2020,Huber2026} and \texttt{aiida-quantumespresso-ph} plugins of the AiiDA framework~\cite{Huber2020,Uhrin2021}.

% =========================================================================================== %
\subsection{Stochastic self-consistent harmonic approximation (SSCHA)}
\label{sec:sschainfo}
% =========================================================================================== %
%
The active-learning SSCHA and SSCHA-DFT calculations reported in Fig.~\ref{fig:alvsdft} for \ce{Li2O} and $\alpha$-\ce{CsPbI3} were run in $2 \times 2 \times 2$ supercells of the respective unit cell, commensurate with the $\mathbf{q}$-point meshes of the DFPT calculations for harmonic dynamical matrices.
For the $\delta$ phase of \ce{CsPbI3}, we ran the on-the-fly active-learning SSCHA calculations in a $1 \times 2 \times 1$ supercell, corresponding to a 40-atom supercell, equivalent to the supercell size used for the $\alpha$ phase.
The calculations are performed by minimizing the SSCHA Gibbs functional~\cite{Monacelli2018}, meaning that the temperature-dependent equilibrium volumes are obtained from variable-cell SSCHA optimizations at fixed pressure $P = 0$, in which the lattice vectors are relaxed using the SCHA stress tensor~\cite{Monacelli2018} simultaneously with the centroids $\bm{R}$ and the auxiliary force constants $\bm{\Phi}$, rather than from a series of fixed-volume free-energy calculations.
We note that the DFPT dynamical matrix of cubic $\alpha$--\ce{CsPbI3} displays imaginary frequencies at the $R$ and $M$ points.
Since the SSCHA trial density matrix requires a positive-definite auxiliary force-constant matrix, the initial guess is made positive definite by reversing the sign of the negative eigenvalues of the mass-rescaled dynamical matrix (e.g., see Ref.~\citenum{Monacelli2021a}, Section 3.2.4).
This concerns the frequencies of the auxiliary force constants $\bm{\Phi}$ only, which are among the parameters optimized by the SSCHA method.
As a remark, these frequencies are different from the phonon frequencies measured in experiments, which are instead obtained as the second-order derivative of the free energy (see Appendix~\ref{sec:hessian}).
At each iteration, we extract a population of 200 and 400 atomic configurations, for \ce{Li2O} and \ce{CsPbI3} respectively, and use a maximum of 20 populations for the SSCHA functional optimization at each temperature, a minimization step of 0.05, a Kong-Liu ratio of 0.5, and a meaningful factor of 0.05, following the nomenclature of the \texttt{python-sscha} code (see Ref.~\citenum{Monacelli2021a} for the definition of these parameters).
For the positional free energy Hessian matrix (Fig.~\ref{fig:alvsdft}(c) and (f)), the free energy is first re-minimized on the final ensemble at the corresponding temperature with a stricter meaningful factor of 0.01, and the Hessian is then evaluated on this ensemble, including anharmonic contributions~\cite{Bianco2017, Monacelli2021a} up to 3$^{\mathrm{rd}}$-order for \ce{Li2O} and 4$^{\mathrm{th}}$-order for $\alpha$-\ce{CsPbI3}.

The production SSCHA free-energy calculations of $\Delta G_{\delta \rightarrow \alpha}$ in Fig.~\ref{fig:phasetransitions} are performed by minimizing the SSCHA Gibbs functional~\cite{Monacelli2018}, as stated above, but in larger supercells containing 2560 atoms and using a single, fixed ML potential, i.e., without further active learning.
As described in the main text, the same potential is employed for both phases, as it is trained on the DFT labels produced by the two on-the-fly active-learning calculations on the $\alpha$ and $\delta$ phases.
In the Supplementary Information (see Supplementary Figure 4 and Supplementary Discussion), we report a careful convergence study of the transition temperature with respect to finite size effects.
We note that supercells of 1080 atoms are already sufficient to converge the transition temperature within a few kelvin, and that supercells of 320 atoms are sufficient to converge the temperature within $\approx 15$~K.
%

% =========================================================================================== %
\subsection{Sparse Gaussian process (SGP)}
% =========================================================================================== %
%
We use the SGP force field implemented in the \textsc{FLARE} code~\cite{Vandermause2020, Vandermause2022}. 
For the particular application shown in this study, we use the $\mathbf{B}^{(2)}$ ACE~\cite{Drautz2019} descriptors with $n_{\mathrm{max}}=8$ and 12 Chebyshev polynomials, respectively for \ce{Li2O} and \ce{CsPbI3}, a maximum angular momentum $l_{\mathrm{max}}=4$, and a radial cutoff of 5.0~\AA.
We use a normalized dot-product kernel with an exponent $\xi=2$. 
These hyperparameters were found to provide good accuracy for $\alpha$--\ce{CsPbI3} on an initial subset of DFT data (see Supplementary Figures 5, 6, 7, and 8).
We found it very important to provide single atomic energies to avoid total energy offsetting during the on-the-fly training, which would be detrimental for the SSCHA minimization. 
We used single-atom energies taken from the ``per-atom-energy'' of the most stable single-element bulk structure in the MC3D database~\cite{Huber2026}.
The model before training is initialized with the following hyperparameters: energy noise $\sigma_E = 0.01~$eV, force noise $\sigma_F = 0.05~$eV/\AA, stress noise $\sigma_s = 10^{-4}~$eV/\AA$^{3}$, and signal variance $\sigma = 2.0$.
Training of the hyperparameters during the on-the-fly active-learning simulation is performed using the L-BFGS-B algorithm with a maximum number of 100 iterations, and using the following (first lower and then upper) bounds for signal variance, energy, force, and stress noise, respectively: $(0.1,6.0),(0.01,1.00),(0.01,1.00),(10^{-4}, 50\times10^{-4})$.
The bounds were used to avoid negative or too large numerical values, and for efficiency reasons with respect to the BFGS optimizer, which would require a larger memory usage, while maintaining comparable accuracy (see Supplementary Figure 9).
%

% =========================================================================================== %
\subsection{On-the-fly active learning}
% =========================================================================================== %
%
The active learning workflow presented in this work is made available in the official \texttt{python-sscha} code as a part of a new class called \texttt{AiiDAEnsemble}~\cite{Cignarella2025}, which also handles the automated submission of DFT jobs.
The on-the-fly calculations used an (SGP) uncertainty threshold of 0.05 to call DFT and 0.005 for the threshold to add the sparse environments (corresponding to 10\% of the uncertainty threshold).
We tried different values of the uncertainty thresholds for structure selection, and found that 0.05 provides a good trade-off between accuracy and efficiency (see Supplementary Figure 9).
Each ensemble was subdivided into batches containing 8 configurations each.
The active-learning simulations are carried out over the entire temperature range shown in Fig.~\ref{fig:alvsdft}, with a temperature step of $\Delta T =50$~K, and using the ``optimal-guess'' strategy (see Appendix~\ref{sec:guessstrategy}).

% ================== REFERENCES
% ========================================================

% ==========================================================
% ============================= SUPPLEMENTARY INFORMATION
% ==========================================================
\clearpage
\onecolumngrid
\setcounter{figure}{0}
\setcounter{table}{0}
\setcounter{equation}{0}
\renewcommand{\figurename}{Supplementary Figure}
\renewcommand{\tablename}{Supplementary Table}
\renewcommand{\thefigure}{\arabic{figure}}
\renewcommand{\thetable}{\Roman{table}}
\renewcommand{\theequation}{\arabic{equation}}
\renewcommand{\theHfigure}{S\arabic{figure}}
\renewcommand{\theHtable}{S\arabic{table}}
\renewcommand{\theHequation}{S\arabic{equation}}

\begin{center}
    {\large\bfseries Supplementary Information for\\[0.3em]
    ``Predicting accurate solid-state thermodynamics is easier than you think''}\\[1em]
    Lorenzo~Bastonero\,\orcidlink{0000-0001-9374-1876}$^{1,*}$,
    Gabriel~Joalland\,\orcidlink{0009-0000-1129-6685}$^{2,3}$,
    Chiara~Cignarella\,\orcidlink{0000-0002-0603-4389}$^{1,2}$,
    Lorenzo~Monacelli\,\orcidlink{0000-0002-6381-3741}$^{2,4}$,
    and Nicola~Marzari\,\orcidlink{0000-0002-9764-0199}$^{1,2,5,6}$\\[0.8em]
    {\small\itshape
    $^{1}$\BREMEN\\
    $^{2}$\THEOSMARVEL\\
    $^{3}$\RENNES\\
    $^{4}$\SAPIENZA\\
    $^{5}$\CAMBRIDGE\\
    $^{6}$\PSI\\[0.5em]}
    {\small $^{*}$\href{mailto:lbastone@uni-bremen.de}{lbastone@uni-bremen.de}}
\end{center}
\vspace{1em}

%% SUPPL. DISCUSSION
%%%%%%%%%%%%%%%%%%%%%%%%%%%%%%%%%%%%%%%%%%%%%%%%%%%%%%%%%%%%%%%%%%%%%
\section*{Supplementary Discussion}
%

% =========================================================================================== %
\subsection*{Quasi-harmonic approximation considerations}
% =========================================================================================== %
The on-the-fly active learning for \ce{Li2O} only required 44 single-point DFT calculations.
The quasi-harmonic approximation requires multiple harmonic phonon calculations at different volumes. 
If we were to perform the harmonic phonon calculations, e.g., using the finite difference approach~\cite{SI-Togo2015,SI-Bastonero2024}, we would need to perform two single-point DFT calculations for each volume, corresponding to the number of irreducible displacements for the \ce{Li2O} Fm$\overline{3}$m crystal.
Therefore, assuming 10 different volumes, the total number of required atomic configurations is about 20, rather close to the amount needed using the proposed on-the-fly scheme.
It is important to note that the results thus obtained through the active-learning approach account for the full anharmonic and quantum nuclear effects, out-of-reach for the QHA method.
Moreover, the generated force-field would be valuable for the calculation of other interesting phonon-related properties, such as thermal transport, and could be used in principle for many other applications.

% =========================================================================================== %
\subsection*{Transition temperature and its uncertainty}
% =========================================================================================== %
The transition temperature between the $\delta$ and $\alpha$ phases of \ce{CsPbI3} corresponds to the temperature at which the Gibbs free energy difference vanishes, $\Delta G_{\delta\rightarrow\alpha}(T_{\delta\leftrightarrow\alpha}) = 0$.
Since the SSCHA free energies are computed at a discrete set of temperatures, we locate this zero through a linear interpolation of the computed free energy differences, $\Delta G_{\delta\rightarrow\alpha}(T) \approx m\,T + q$, so that $T_{\delta\leftrightarrow\alpha} = -q/m$.
The coefficients $m$ and $q$ are determined by a weighted linear least-squares fit through all computed temperatures, in which each point is weighted by $1/\sigma_i^2$, where $\sigma_i$ is the stochastic error of $\Delta G_{\delta\rightarrow\alpha}(T_i)$.
The latter is obtained from the stochastic errors that SSCHA computes for the free energy of each phase~\cite{SI-Monacelli2021a} -- reflecting the finite size of the ensemble used to evaluate the free energy -- normalized per atom and summed in quadrature.
The uncertainty $\sigma_{T}$ on the transition temperature is then propagated from the covariance matrix of the fit, treating the SSCHA error bars as absolute:
\begin{equation}
    \sigma_{T}^2 = \frac{\mathrm{var}(q)}{m^2} + \frac{q^2}{m^4}\,\mathrm{var}(m) - 2\,\frac{q}{m^3}\,\mathrm{cov}(q,m).
\end{equation}
We note that $\Delta G_{\delta\rightarrow\alpha}(T)$ displays a small curvature, as expected from a temperature-dependent entropy difference, so that a straight line can slightly underfit the data.
We quantify the quality of the fit through the reduced chi-square $\chi^2/\mathrm{dof}$, where $\chi^2 = \sum_i \left[ \Delta G_{\delta\rightarrow\alpha}(T_i) - (m\,T_i + q) \right]^2 / \sigma_i^2$ is the weighted sum of the squared residuals and $\mathrm{dof}$ is the number of degrees of freedom of the fit, i.e., the number of fitted temperatures minus the two free parameters $m$ and $q$.
Whenever $\chi^2/\mathrm{dof} > 1$ -- indicating that the residuals exceed what the stochastic error bars alone can account for -- we conservatively rescale $\sigma_{T}$ by $\sqrt{\chi^2/\mathrm{dof}}$~\cite{SI-Birge1932} to restore consistency between the quoted uncertainty and the observed scatter of the data around the model.
The transition temperatures obtained in this way, for all supercell and ensemble sizes considered, are collected in Supplementary Table~\ref{tab:tc_convergence}.

% =========================================================================================== %
\subsection*{Free-energy interpolation to larger supercells}
% =========================================================================================== %
The Gibbs free energy differences governing the $\delta \leftrightarrow \alpha$ transition are of the order of a few meV/atom, so that even small finite-size errors can shift the predicted transition temperature by tens of kelvin (see Supplementary Table~\ref{tab:tc_convergence}), while the cost of an explicit SSCHA calculation grows rapidly with the supercell size.
It is therefore important to verify that the transition temperature is converged with respect to the supercell size beyond what can be explicitly afforded; to this end, we reconstruct the free energy of larger supercells from the converged calculations on smaller ones at negligible computational cost, validating the reconstruction against the explicitly computed 2560-atom results.
The SSCHA free energy can be decomposed as
\begin{equation}
    F[\bm{\Phi}]
    = \langle V \rangle_{\rho_{\bm{\Phi}}} + \left( \langle K \rangle_{\rho_{\bm{\Phi}}} - T S[\rho_{\bm{\Phi}}] \right)
    = \mathcal{F}[\bm{\Phi}] + \langle V - E \rangle_{\rho_{\bm{\Phi}}},
    \label{eq:fe_decomposition}
\end{equation}
where $\rho_{\bm{\Phi}}$ is the SSCHA density matrix, $\langle V \rangle$ and $\langle K \rangle$ are the average potential and kinetic energy, $S$ is the entropy, $E$ is the potential energy of the auxiliary harmonic Hamiltonian defined by the force constants $\bm{\Phi}$, and $\mathcal{F}[\bm{\Phi}] = \langle E \rangle_{\rho_{\bm{\Phi}}} + \langle K \rangle_{\rho_{\bm{\Phi}}} - T S[\rho_{\bm{\Phi}}]$ is its harmonic free energy.
The anharmonic term $\langle V - E \rangle$ converges much faster with the supercell size than the harmonic free energy $\mathcal{F}$, so that the residual finite-size error is dominated by the latter -- which, in turn, can be evaluated on denser Brillouin-zone samplings at negligible cost through Fourier interpolation of the auxiliary force constants.
Denoting by $\mathcal{F}_{N}$ the harmonic free energy per atom of the auxiliary force constants $\bm{\Phi}_{N}$ of the converged $N$-atom calculation, and by $\mathcal{F}_{N \rightarrow N'}$ the same quantity after Fourier interpolation of $\bm{\Phi}_{N}$ onto the $\mathbf{q}$-point mesh commensurate with the larger supercell containing $N' > N$ atoms, the free energy per atom of the latter can be reconstructed as
\begin{equation}
    F_{N'} \approx F_{N} + \mathcal{F}_{N \rightarrow N'} - \mathcal{F}_{N},
    \label{eq:fe_interpolation}
\end{equation}
i.e., the harmonic free energy of the small supercell is replaced by its interpolated counterpart, while the fast-converging anharmonic average $\langle V - E \rangle$ is kept at its $N$-atom value.
Since this correction is a deterministic function of the auxiliary force constants, the interpolated free energy inherits the stochastic error of the source calculation.
Applying Supplementary Equation~\eqref{eq:fe_interpolation} to both phases at each temperature yields the interpolated transition temperatures reported in Supplementary Table~\ref{tab:tc_convergence}: the value reconstructed from 1080 to 2560 atoms reproduces the explicitly computed 2560-atom result within $1\sigma$ for both ensemble sizes, and the two independent reconstructions with 5000 atoms (from the 1080- and from the 2560-atom calculations) agree with each other and with the 2560-atom results to within a few kelvin, showing that the transition temperature is converged with respect to the supercell size.
We finally note that non-analytical corrections~\cite{SI-Gonze1997} to the dynamical matrices are not taken into account in this procedure, since the machine-learned potential is short-ranged and its auxiliary force constants therefore do not carry the long-range dipole--dipole tail.

%% SUPPLEMENTARY FIGURES
%%%%%%%%%%%%%%%%%%%%%%%%%%%%%%%%%%%%%%%%%%%%%%%%%%%%%%%%%%%%%%%%%%%%%
\clearpage
\section*{Supplementary Figures}

% --------------------
\begin{figure}[h!]
    \centering
    \includegraphics[width=0.5\textwidth]{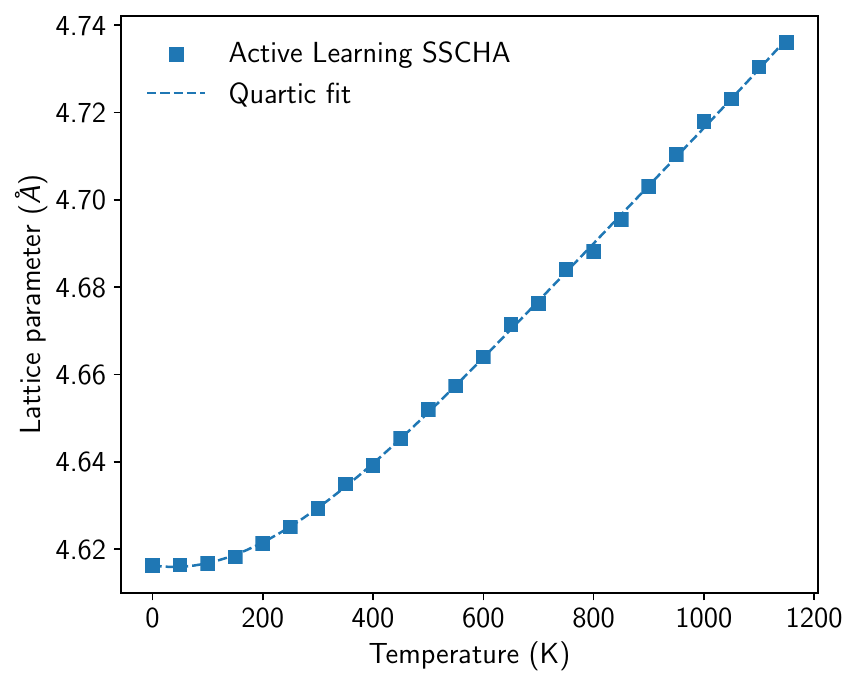}
    \caption{
        Lattice parameter $a$ as obtained from the active learning SSCHA calculations compared to the fourth-order polynomial fit ($a_T = p_0+p_1T+p_2T^2+p_3T^3+p_4T^4$). The polynomial coefficients $p_i$ were fitted using the \texttt{polyfit} routine of the \texttt{numpy} library. 
    }
    \label{fig:linear_expansion}
\end{figure}
% --------------------
\begin{figure}[h!]
    \centering
    \includegraphics[width=0.8\textwidth]{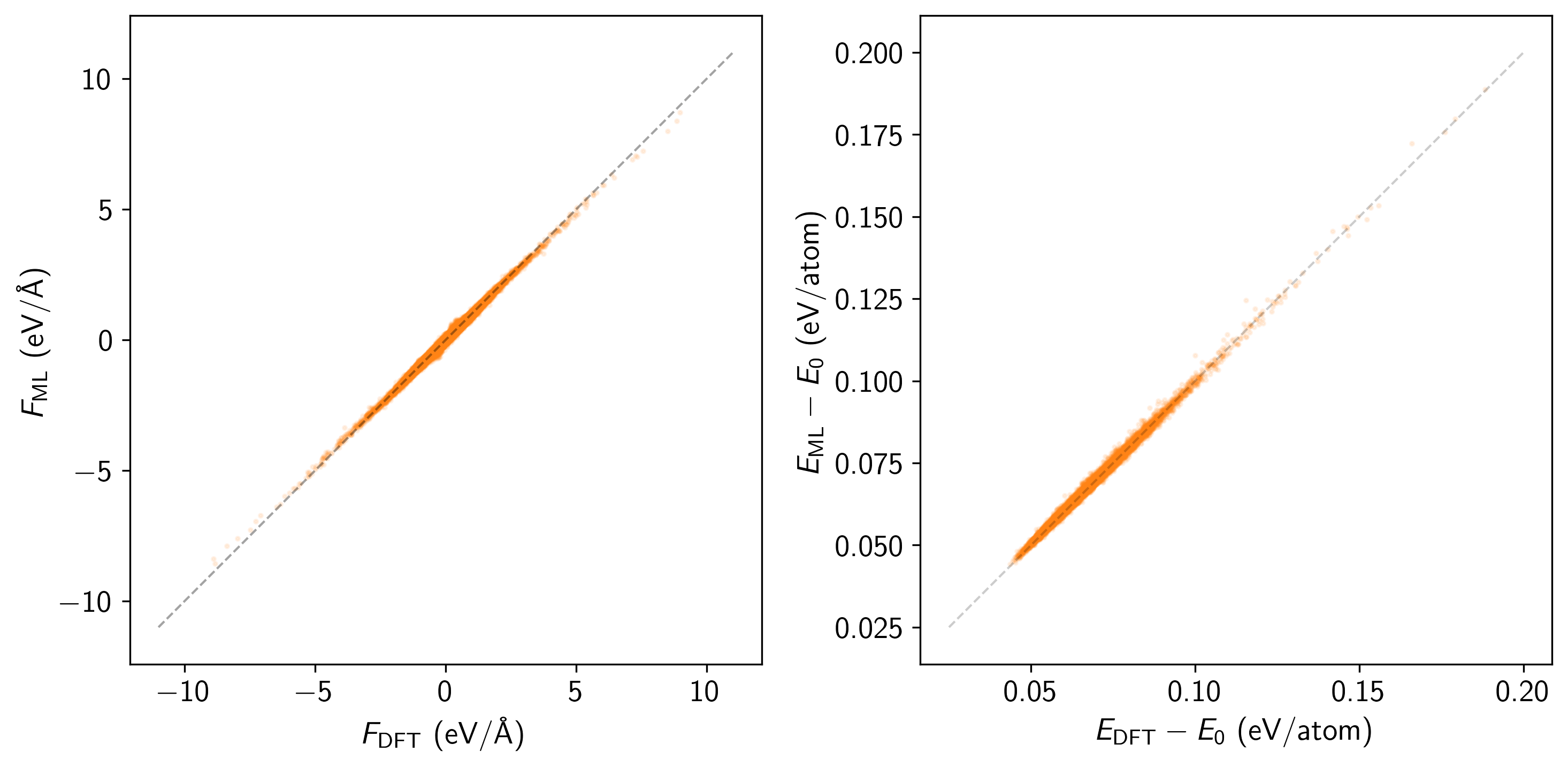}
    \caption{Force components parity plot for the $\alpha$--\ce{CsPbI3} SSCHA--DFT dataset between the final SGP model ($F_{\mathrm{ML}}$), trained on the screened dataset composed of 172 structures from both $\alpha$ and $\delta$ phase (see also main text), and the DFT reference ($F_{\mathrm{DFT}}$).}
    \label{fig:parity_plot_cspbi3}
\end{figure}
% --------------------
\begin{figure}[h!]
    \centering
    \includegraphics[width=0.8\textwidth]{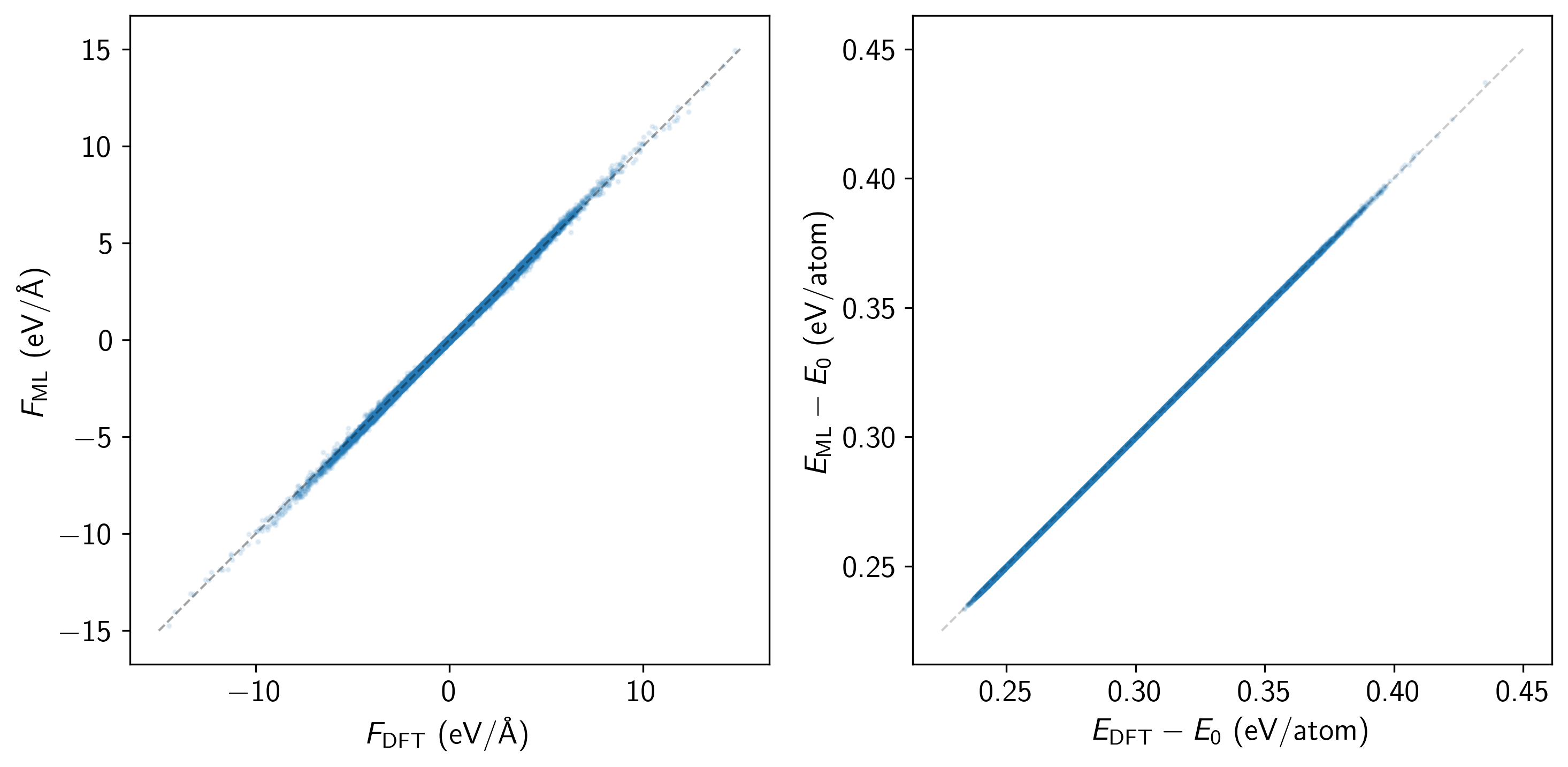}
    \caption{Force components parity plot for the \ce{Li2O} SSCHA--DFT dataset between the final SGP model ($F_{\mathrm{ML}}$) and the DFT reference ($F_{\mathrm{DFT}}$).}
    \label{fig:parity_plot_li2o}
\end{figure}
% --------------------
\begin{figure}[h!]
    \centering
    \includegraphics[width=0.98\textwidth]{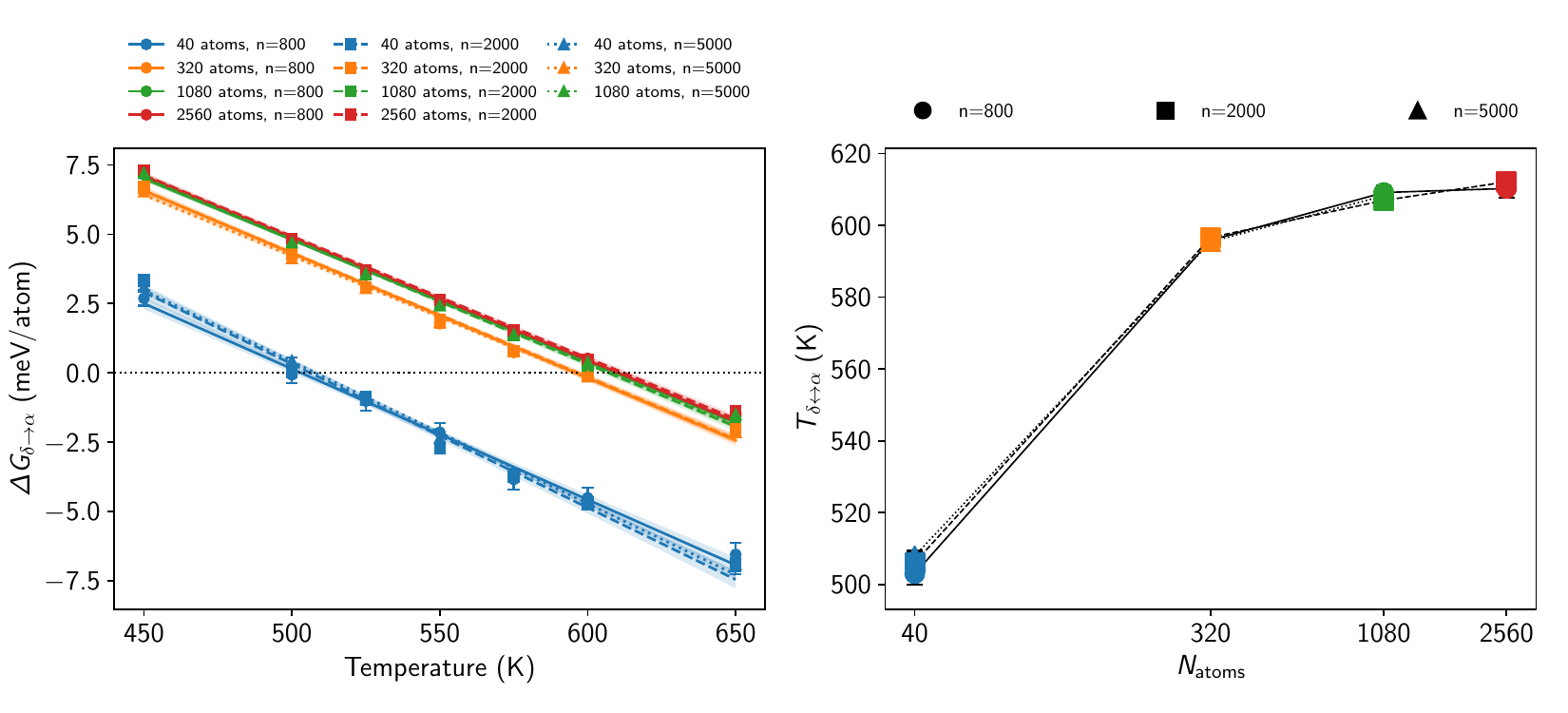}
    \caption{Supercell size convergence, in terms of the number of atoms $N_{\mathrm{atoms}}$ in the supercell, of the $\delta$ to $\alpha$ phase (left panel) Gibbs free energy difference $\Delta G_{\delta\rightarrow\alpha}$ and (right panel) transition temperature $T_{\delta\leftrightarrow\alpha}$ for \ce{CsPbI3} using SSCHA with the ML potential, for ensembles of $N_{\mathrm{configs}} = 800$, $2000$, and $5000$ configurations (color encodes the supercell size; marker and line style the ensemble size).
    In the left panel, the error bars are the SSCHA stochastic errors, and the lines are the $1/\sigma^2$-weighted linear fits whose zero defines the transition temperature, shown with their $1\sigma$ confidence bands (see Supplementary Discussion).
    In the right panel, the error bars are the fit uncertainties $\sigma_{T}$ (see Supplementary Discussion).
    }
    \label{fig:supercell_convergence_cspbi3}
\end{figure}
% --------------------
\begin{figure}[h!]
    \centering
    \includegraphics[width=0.7\textwidth]{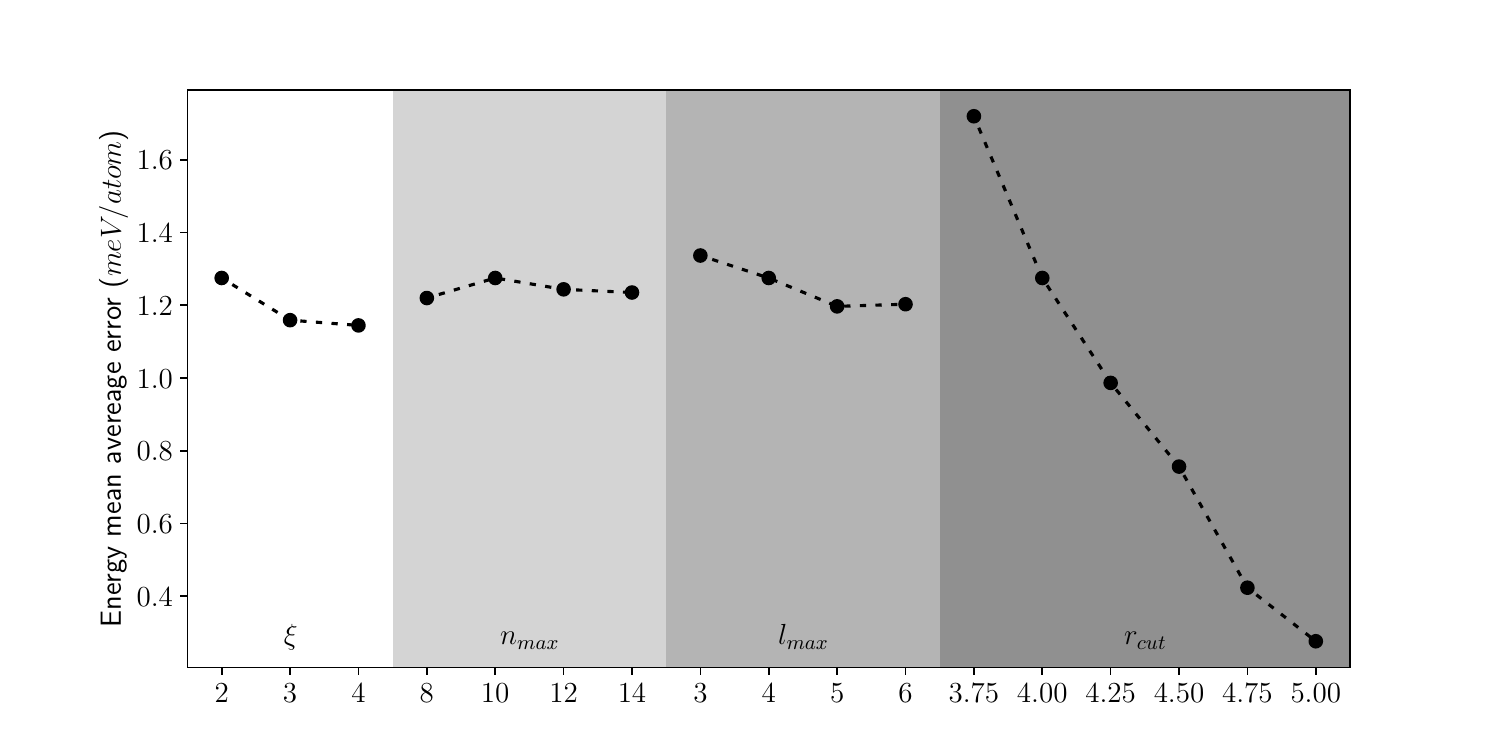}
    \caption{The Sparse Gaussian processes mean average errors on energies for various hyperparameters of the kernel and the ACE descriptor. These are evaluated for $\alpha$--\ce{CsPbI3} on a subset of the first-principles SSCHA dataset.}
    \label{fig:sgp_energy_test}
\end{figure}
% --------------------
\begin{figure}[h!]
    \centering
    \includegraphics[width=0.7\textwidth]{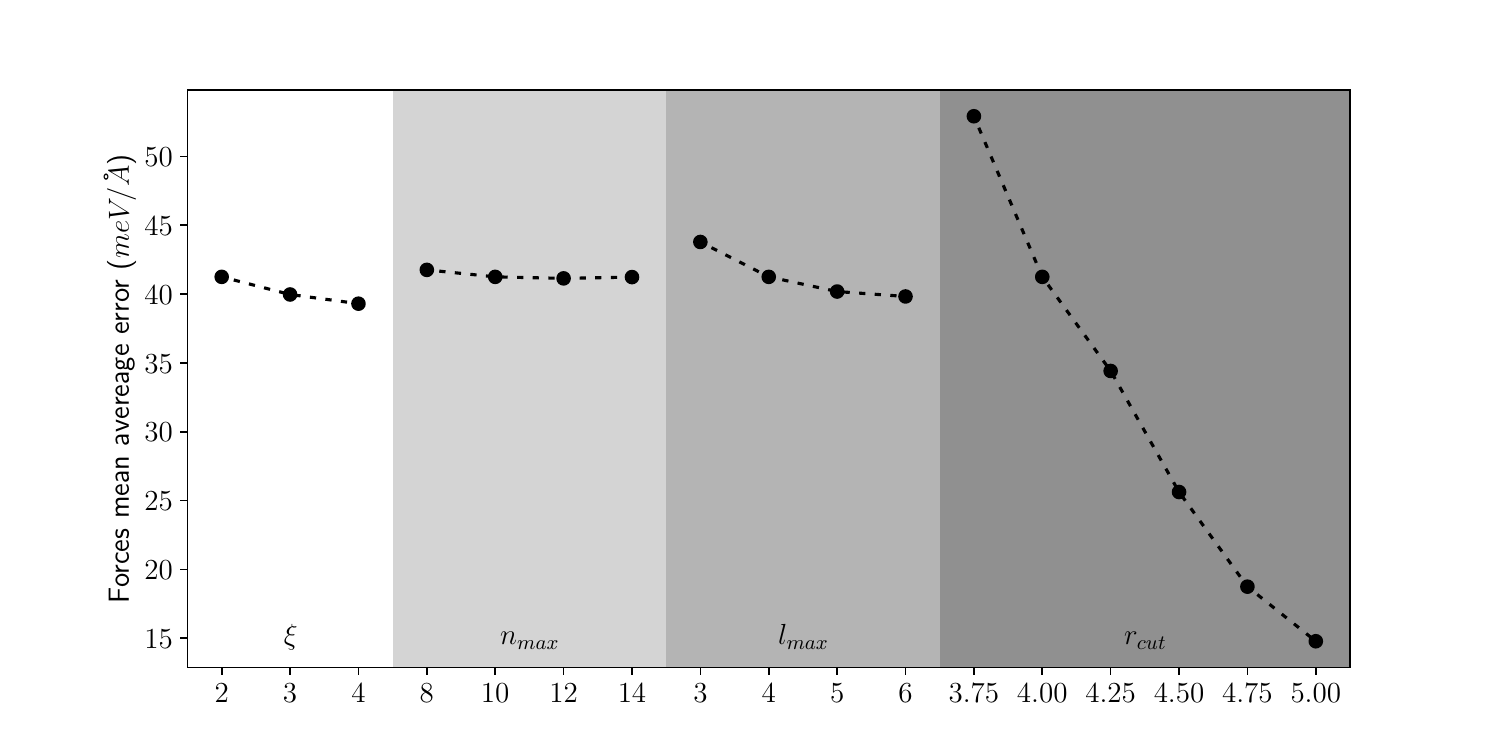}
    \caption{The Sparse Gaussian processes mean average errors on forces for various hyperparameters of the kernel and the ACE descriptor. These are evaluated for $\alpha$--\ce{CsPbI3} on a subset of the first-principles SSCHA dataset.}
    \label{fig:sgp_forces_test}
\end{figure}
% --------------------
\begin{figure}[h!]
    \centering
    \includegraphics[width=0.7\textwidth]{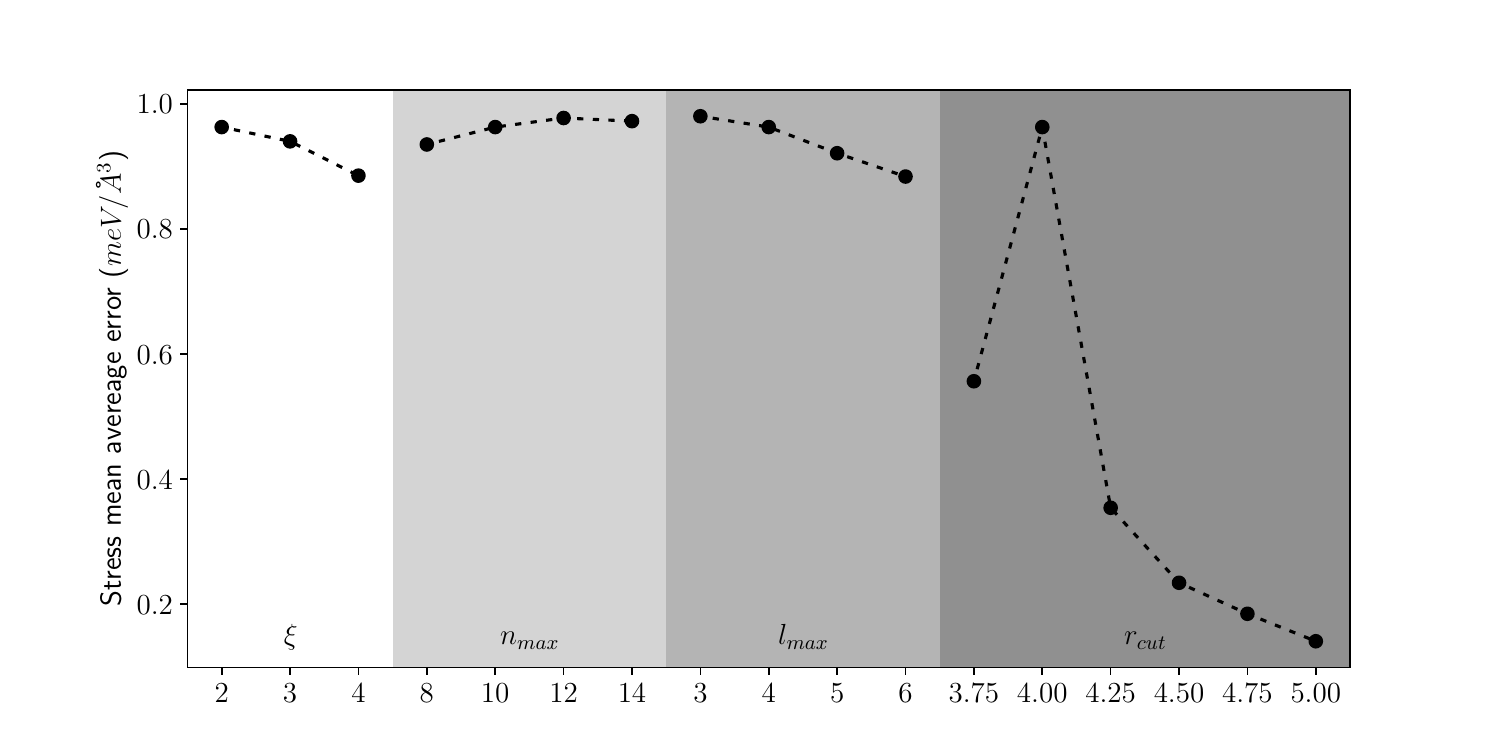}
    \caption{The Sparse Gaussian processes mean average errors on stresses for various hyperparameters of the kernel and the ACE descriptor. These are evaluated for $\alpha$--\ce{CsPbI3} on a subset of the first-principles SSCHA dataset.}
    \label{fig:sgp_stress_test}
\end{figure}
% --------------------
\begin{figure}[h!]
    \centering
    \includegraphics[width=0.7\textwidth]{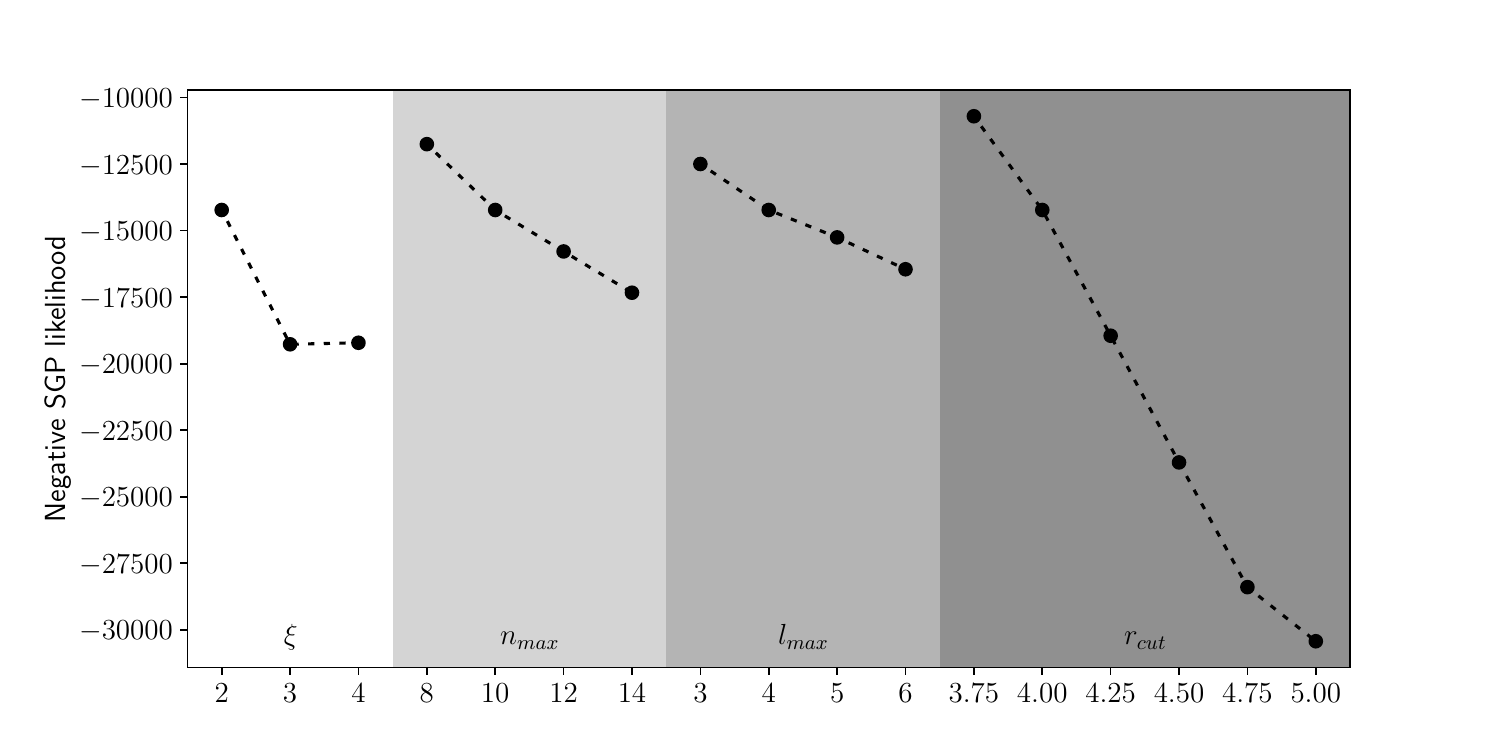}
    \caption{The Sparse Gaussian processes likelihood for various hyperparameters of the kernel and the ACE descriptor. These are evaluated for $\alpha$--\ce{CsPbI3} on a subset of the first-principles SSCHA dataset.}
    \label{fig:sgp_likelihood_test}
\end{figure}
% --------------------
\begin{figure}[h!]
    \centering
    \includegraphics[width=0.5\textwidth]{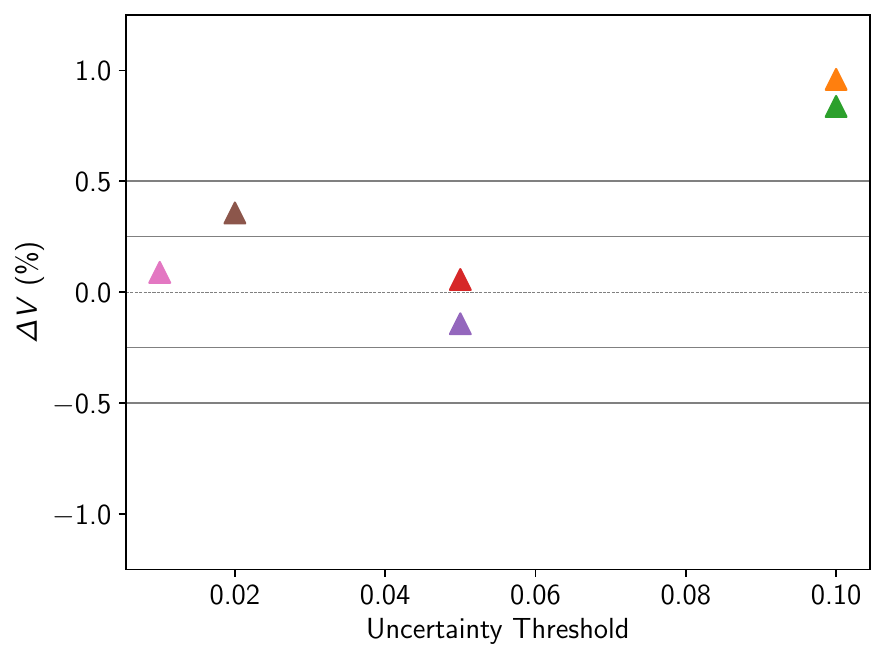}
    \caption{
        Volume difference in percentage at 250~K for $\alpha$--\ce{CsPbI3}, $\Delta V = 100\cdot(V_{\mathrm{AL}}/V_{\mathrm{ref}}-1)$, between the SSCHA-DFT reference ($V_{\mathrm{ref}}$) and the SSCHA active learning ($V_{\mathrm{AL}}$) as a function of the uncertainty threshold chosen for selecting structures for DFT labeling.
        For two uncertainty thresholds (0.10 and 0.05), we report two results stemming from active-learning calculations carried out with different optimizers (BFGS and L-BFGS-B). 
    }
    \label{fig:sgp_errors}
\end{figure}
% --------------------

%% SUPPLEMENTARY TABLES
%%%%%%%%%%%%%%%%%%%%%%%%%%%%%%%%%%%%%%%%%%%%%%%%%%%%%%%%%%%%%%%%%%%%%
\clearpage
\section*{Supplementary Tables}

% --------------------
\begin{table}[h!]
    \centering
    \begin{tabular}{lccccc}
    \toprule
    Dataset & $N_{\mathrm{train}}$ &$N_{\mathrm{config}}$ & $E_{\mathrm{MAE}}$ & $F_{\mathrm{MAE}}$ & $\sigma_{\mathrm{MAE}}$  \\
    {} & {} & {} & (meV/atom) & (meV/\AA) & (meV/\AA$^{3}$ $|$ kbar) \\
    \midrule
    $\alpha$--\ce{CsPbI3} & 172 & 16\,000 & 0.67 & 23.7 & 0.20 $|$ 0.3 \\
    \ce{Li2O} & 44 & 21\,347 & 0.20 & 11.3 & 0.23 $|$ 0.4 \\
    \bottomrule
    \end{tabular}
    \caption{
    Mean absolute error (MAE) statistics (energy $E$, forces $F$, and stress $\sigma$) of the final ML potentials, as obtained by training on the dataset generated during the on-the-fly active learning comprising a small number of training structures ($N_{\mathrm{train}}$), against the reference SSCHA-DFT datasets comprising thousands of configurations ($N_{\mathrm{config}}$).
    The ML potential for \ce{CsPbI3} corresponds to the one trained on the screened dataset composed of 172 structures from both $\alpha$ and $\delta$ phase (see also main text).
    }
    \label{tab:placeholder}
\end{table}
% --------------------
\begin{table}[h!]
    \centering
    \begin{tabular}{lccc}
    \toprule
    {} & \multicolumn{3}{c}{$T_{\delta\leftrightarrow\alpha}$ (K)} \\
    \cmidrule(lr){2-4}
    $N_{\mathrm{atoms}}$ & $N_{\mathrm{configs}} = 800$ & $N_{\mathrm{configs}} = 2000$ & $N_{\mathrm{configs}} = 5000$ \\
    \midrule
    40   & $503.0 \pm 3.0$ & $506.1 \pm 3.2$ & $508.0 \pm 1.6$ \\
    320  & $595.8 \pm 2.8$ & $596.6 \pm 2.5$ & $595.4 \pm 2.0$ \\
    1080 & $609.1 \pm 2.0$ & $606.9 \pm 2.6$ & $608.2 \pm 2.7$ \\
    2560 & $610.2 \pm 2.6$ & $612.1 \pm 2.5$ & -- \\
    \midrule
    $1080 \rightarrow 2560$ & $612.7 \pm 2.1$ & $610.4 \pm 2.7$ & -- \\
    $1080 \rightarrow 5000$ & $613.9 \pm 2.1$ & $611.5 \pm 2.7$ & -- \\
    $2560 \rightarrow 5000$ & $611.4 \pm 2.6$ & $613.3 \pm 2.5$ & -- \\
    \bottomrule
    \end{tabular}
    \caption{
    Transition temperature $T_{\delta\leftrightarrow\alpha}$ of \ce{CsPbI3} as a function of the number of atoms $N_{\mathrm{atoms}}$ in the supercell and of the number of configurations $N_{\mathrm{configs}}$ in the SSCHA ensemble.
    Each value is the zero of a $1/\sigma^2$-weighted linear fit of $\Delta G_{\delta\rightarrow\alpha}(T)$, and the quoted uncertainty is propagated from the fit covariance (see Supplementary Discussion).
    Rows labelled $N \rightarrow N'$ are not explicit calculations with $N'$ atoms: they are reconstructed from the $N$-atom calculations by interpolating the auxiliary force constants to the supercell with $N'$ atoms (see Supplementary Discussion); the reported transition temperature therefore refers to the target size $N'$.
    The experimental transition temperature is $T_{\mathrm{exp}} \simeq 593$~K~\cite{SI-Ke2021}.
    }
    \label{tab:tc_convergence}
\end{table}
% --------------------

%% BIBLIOGRAPHY
%%%%%%%%%%%%%%%%%%%%%%%%%%%%%%%%%%%%%%%%%%%%%%%%%%%%%%%%%%%%%%%%%%%%%
\clearpage

\clearpage
\renewcommand{\bibsection}{\section*{Supplementary References}}

\end{document}